%% file: latency.tex
\documentclass[journal,10pt]{IEEEtran}

\newtheorem{theorem}{Theorem}

\newtheorem{corollary}{Corollary}
\newtheorem{lemma}{Lemma}

\newtheorem{definition}{Definition}

\newcommand{\be}{\begin{equation}}
\newcommand{\ee}{\end{equation}}
\newcommand{\bea}{\begin{eqnarray}}
\newcommand{\eea}{\end{eqnarray}}
\newcommand{\bw}{\begin{eqnarray*}}
\newcommand{\ew}{\end{eqnarray*}}

\newcommand{\D}{\displaystyle}

\usepackage{amssymb,amsmath}
\usepackage{amsmath}
\usepackage{latexsym}
\usepackage{graphicx}
\usepackage{graphics}
\usepackage{color}
\usepackage{citesort}

\input{psfig.sty}
\usepackage{psfig}

\renewcommand{\baselinestretch}{0.96}

\begin{document}

\title{Joint Latency and Cost Optimization for Erasure-coded Data Center Storage }


\author{Yu Xiang, Tian Lan, Vaneet Aggarwal,  and Yih-Farn R. Chen\thanks{Y. Xiang and T. Lan are with Department of ECE, George Washington University, DC 20052 (email: \{xy336699, tlan\}@gwu.edu). V. Aggarwal and Y. R. Chen are with AT$\&$T Labs-Research, Bedminster, NJ 07921 (email: \{vaneet, chen\}@research.att.com). This work was presented in part at the IFIP Performance, Oct. 2014.}
}

\maketitle

\begin{abstract}
Modern distributed storage systems offer large capacity to satisfy the exponentially increasing need of storage space. They often use erasure codes to protect against disk and node failures to increase reliability, while trying to meet the latency requirements of the applications and clients. This paper provides an insightful upper bound on the average service delay of such erasure-coded storage with arbitrary service time distribution and consisting of multiple heterogeneous files.
Not only does the result supersede known delay bounds that only work for a single file or homogeneous files, it also enables a novel problem of joint latency and storage cost minimization over three dimensions: selecting the erasure code, placement of encoded chunks, and optimizing scheduling policy. The problem is efficiently solved via the computation of a sequence of convex approximations with provable convergence. We further prototype our solution in an open-source, cloud storage deployment over three geographically distributed data centers. Experimental results validate our theoretical delay analysis and show significant latency reduction, providing valuable insights into the proposed latency-cost tradeoff in erasure-coded storage.
\end{abstract}



\vspace{-.2in}

\input{1_intro}

\vspace{-.2in}

\input{3_model_v2}

\vspace{-.2in}

\input{4_bound_v2}
\vspace{-.2in}

\input{5_algo_v2}
\vspace{-.2in}
\input{6_impl_latest}

\vspace{-.2in}

\section{Conclusions}
\vspace{-.1in}

Relying on a novel probabilistic scheduling policy, this paper develops an analytical upper bound on average service delay of erasure-coded storage with arbitrary number of files and any service time distribution. A joint latency and cost minimization is formulated by collectively optimizing over erasure code, chunk placement, and scheduling policy. The minimization is solved using an efficient algorithm with proven convergence. Even though only local optimality can be guaranteed due to the non-convex nature of the mixed-integer optimization problem, the proposed algorithm significantly reduces a latency-plus-cost objective. Both our theoretical analysis and algorithm design are validated via a prototype in Tahoe, an open-source, distributed file system. Several practical design issues in erasure-coded, distributed storage, such as incorporating network latency and dynamic data management have been ignored in this paper and open up avenues for future work.


\vspace{-.2in}
\appendix

\vspace{-.15in}

\subsection{Proof of Theorem \ref{th:thm_prob}}
We first prove that the conditions $\sum_{j=1}^m \pi_{i,j} = k_i$ $\forall i$ and $\pi_{i,j}\in [0,1]$ are necessary. $\pi_{i,j}\in [0,1]$ for all $i,j$ is obvious due to its definition. Then, it is easy to show that
\begin{eqnarray}
& \D \sum_{j=1}^m \pi_{i,j} & = \sum_{j=1}^m \sum_{\mathcal{A}_i\subseteq \mathcal{S}_i,j\in\mathcal{A}_i} \mathbb{P}(\mathcal{A}_i) \nonumber \\
&  & =\sum_{\mathcal{A}_i\subseteq \mathcal{S}_i} \sum_{j\in\mathcal{A}_i}  \mathbb{P}(\mathcal{A}_i) \nonumber \\
& & =\sum_{\mathcal{A}_i\subseteq \mathcal{S}_i}  k_i \mathbb{P}(\mathcal{A}_i)  = k_i \label{eq:k_1}
\end{eqnarray}
where the first step is due to  (\ref{eq:pi}), the second step changes the order of summation, the last step uses the fact that each set $\mathcal{A}_i$ contain exactly $k_i$ nodes and that $\sum_{\mathcal{A}_i\subseteq \mathcal{S}_i} \mathbb{P}(\mathcal{A}_i)=1$.

Next, we prove that for any set of $\pi_{i,1},\ldots,\pi_{i,m}$ (i.e., node selection probabilities of file $i$) satisfying $\sum_{j=1}^m \pi_{i,j} = k_i$ and $\pi_{i,j}\in [0,1]$, there exists a probabilistic scheduling scheme with feasible load balancing probabilities $\mathbb{P}(\mathcal{A}_i)$ $\forall \mathcal{A}_i \subseteq \mathcal{S}_i$ to achieve the same node selection probabilities. We start by constructing $\mathcal{S}_i=\{j: \pi_{i,j} >0\}$, which is a set containing at least $k_i$ nodes, because there must be at least $k_i$ positive probabilities $\pi_{i,j}$ to satisfy $\sum_{j=1}^m \pi_{i,j} = k_i$. Then, we choose erasure code length $n_i=|\mathcal{S}_i|$ and place chunks on nodes in  $\mathcal{S}_i$. From (\ref{eq:pi}), we only need to show that when $\sum_{j\in\mathcal{S}_i} \pi_{i,j} = k_i$ and $\pi_{i,j}\in [0,1]$, the following system of $n_i$ linear equations have a feasible solution $\mathbb{P}(\mathcal{A}_i)$ $\forall \mathcal{A}_i\subseteq \mathcal{S}_i$:
\begin{eqnarray}
\sum_{\mathcal{A}_i\subseteq \mathcal{S}_i} {\bf 1}_{\{ j\in\mathcal{A}_i\}} \cdot \mathbb{P}(\mathcal{A}_i)=\pi_{i,j}, \ \forall j\in \mathcal{S}_i \label{eq:pi_1}
\end{eqnarray}
where ${\bf 1}_{\{ j\in\mathcal{A}_i\}}$ is an indicator function, which is 1 if $j\in\mathcal{A}_i$, and 0 otherwise. We will make use of the following lemma.

\begin{lemma} Farkas-Minkowski Theorem \cite{Angell:02}. Let ${\bf A}$ be an $m\times n$ matrix with real entries, and ${\bf x}\in \mathbb{R}^{n}$ and ${\bf b}\in \mathbb{R}^{m}$ be 2 vectors. A necessary and sufficient condition that ${\bf A}\cdot{\bf x}={\bf b}, \ {\bf x}\ge 0$ has a solution is that, for all ${\bf y}\in \mathbb{R}^m$ with the property that ${\bf A}^T \cdot {\bf y}\ge 0$, we have  $\langle{\bf y}, {\bf b}\rangle \ge 0$.
\end{lemma}

We prove the desired result using mathematical induction. It is easy to show that the statement holds for $n_i=k_i$. In this case, we have a unique solution $\mathcal{A}_i = \mathcal{S}_i$ and $\mathbb{P}(\mathcal{A}_i)=\pi_{i,j}=1$ for the system of linear equations (\ref{eq:pi_1}), because all chunks must be selected to recover file $i$. Now assume that the system of linear equations (\ref{eq:pi_1}) has a feasible solution for some $n_i\ge k_i$. Consider the case with arbitrary  $|\mathcal{S}_i+\{h\}|=n_i+1$ and $\pi_{i,h}+\sum_{j\in\mathcal{S}_i} \pi_{i,j}= k_i$. We have a system of linear equations:
\begin{eqnarray}
\sum_{\mathcal{A}_i\subseteq \mathcal{S}_i+\{h\}} {\bf 1}_{\{ j\in\mathcal{A}_i\}} \cdot \mathbb{P}(\mathcal{A}_i)=\pi_{i,j}, \ \forall j\in \mathcal{S}_i+\{h\} \label{eq:pi_2}
\end{eqnarray}
Using the Farkas-Minkowski Theorem \cite{Angell:02}, a sufficient and necessary condition that (\ref{eq:pi_2}) has a non-negative solution is that, for any $y_1,\ldots, y_m$ and $\sum_{j} y_j \pi_{i,j}<0$, we have
\begin{eqnarray}
\sum_{j\in \mathcal{S}_i+\{h\}} y_j {\bf 1}_{\{ j\in\mathcal{A}_i\}}  < 0 \ {\rm for \ some} \ \mathcal{A}_i\subseteq \mathcal{S}_i+\{h\}.\label{eq:y_1}
\end{eqnarray}

Toward this end, we construct $\hat{\pi}_{i,j} =\pi_{i,j} + [u-\pi_{i,j}]^{+}$ for all $j\in \mathcal{S}_i$. Here $[x]^{+}=\max(x, 0)$ is a truncating function and $u$ is a proper {\em water-filling level} satisfying
\begin{eqnarray}
\sum_{j\in \mathcal{S}_i}  [u-\pi_{i,j}]^{+} = \pi_{i,h}.  \label{eq:h_1}
\end{eqnarray}
It is easy to show that $\sum_{j\in \mathcal{S}_i} \hat{\pi}_{i,j}= \pi_{i,h} +\sum_{j\in \mathcal{S}_i} \pi_{i,j} =k_i$ and $\hat{\pi}_{i,j}\in[0,1]$, because $\hat{\pi}_{i,j} = \max(u,{\pi}_{i,j} ) \in[0, 1]$. Here we used the fact that $u<1$ since $k_i=\sum_{j\in \mathcal{S}_i} \hat{\pi}_{i,j}\ge\sum_{j\in \mathcal{S}_i} u \ge k_iu  $.  Therefore, the system of linear equations in (\ref{eq:pi_1}) with $\hat{\pi}_{i,j}$ on the right hand side must have a non-negative solution due to our induction assumption for $n_i=|\mathcal{S}_i|$. Furthermore, without loss of generality, we assume that $y_h\ge y_j$ for all $j\in \mathcal{S}_i$ (otherwise a different $h$ can be chosen). It implies that
\begin{eqnarray}
& \D \sum_{j\in \mathcal{S}_i} y_j \hat{\pi}_{i,j} & =  \sum_{j\in \mathcal{S}_i} y_j (\pi_{i,j}+[u-\pi_{i,j}]^{+}) \nonumber \\
& & \le \sum_{j\in \mathcal{S}_i} y_j \pi_{i,j} + \sum_{j\in \mathcal{S}_i} y_h \pi_{i,j} \nonumber \\
& & = \sum_{j\in \mathcal{S}_i} y_j \pi_{i,j} + y_h \pi_{i,h} <0, \label{eq:h_2}
\end{eqnarray}
where the second step follows from (\ref{eq:h_1}) and the last step uses $\sum_{j} y_j \pi_{i,j}<0$.

Applying the Farkas-Minkowski Theorem to the system of linear equations in (\ref{eq:pi_1}) with $\hat{\pi}_{i,j}$ on the right hand side, the existence of  a non-negative solution (due to our induction assumption for $n_i$) implies that $\sum_{j\in \mathcal{S}_i} y_j {\bf 1}_{\{ j\in\mathcal{A}_i\}}  < 0$ for some $\hat{\mathcal{A}_i}\subseteq \mathcal{S}_i$. It means that
\begin{eqnarray}
\sum_{j\in \mathcal{S}_i+\{h\}} y_j {\bf 1}_{\{ j\in\hat{\mathcal{A}}_i\}}  = y_h {\bf 1}_{\{ h\in\hat{\mathcal{A}}_i\}}+ \sum_{j\in \mathcal{S}_i} y_j {\bf 1}_{\{ j\in\hat{\mathcal{A}}_i\}} <0.
\end{eqnarray}
The last step uses ${\bf 1}_{\{ h\in\hat{\mathcal{A}}_i\}}=0$ since $h\notin\mathcal{S}_i$ and $\hat{\mathcal{A}_i}\subseteq \mathcal{S}_i$. This is exactly the desired inequality in (\ref{eq:y_1}). Thus, (\ref{eq:pi_2}) has a non-negative solution due to the Farkas-Minkowski Theorem. The induction statement holds for $n_i+1$. Finally, the solution indeed gives a probability distribution since $\sum_{\mathcal{A}_i\subseteq \mathcal{S}_i+\{h\}}  \mathbb{P}(\mathcal{A}_i) =\sum_j \pi_{i,j} /k_i=1$ due to (\ref{eq:k_1}). This completes the proof.
$\square$
\vspace{-.15in}

%
%

\subsection{Proof of Lemma \ref{th:lemma_1}}
\noindent {\em Proof:} Let ${\bf Q}_{max}$ be the maximum of waiting time $\{{\bf Q}_j, j\in \mathcal{A}_i\}  $. We first show that ${\bf Q}_{max}$ is upper bounded by the following inequality for arbitrary $z\in\mathbb{R}$:
\begin{eqnarray}
{\bf Q}_{max} \le z +\left[{\bf Q}_{\max} - z\right]^{+}  \le z + \sum_{j\in\mathcal{A}_i} \left[{\bf Q}_j - z\right]^{+}, \label{eq:lemma1_p1}
\end{eqnarray}
where $[a]^{+}=\max\{ a, 0\}$ is a truncate function. Now, taking the expectation on both sides of (\ref{eq:lemma1_p1}), we have
\begin{eqnarray}
& \mathbb{E}\left[{\bf Q}_{max} \right] & \le z +  \mathbb{E}\left[ \sum_{j\in\mathcal{A}_i} \left[{\bf Q}_j - z\right]^{+} \right] \nonumber \\
& & = z +  \mathbb{E}\left[ \sum_{j\in\mathcal{A}_i} \frac{1}{2} ( {\bf Q}_j - z + |{\bf Q}_j - z|) \right] \nonumber \\
& & =  z +    \mathbb{E}_{\mathcal{A}_i}\left[ \sum_{j\in\mathcal{A}_i} \frac{1}{2} ( \mathbb{E} [{\bf Q}_j] - z +  \mathbb{E}  |{\bf Q}_j - z|) \right], \nonumber \\
& &  =  z +  \sum_{j\in\mathcal{A}_i} \frac{\pi_{i,j}}{2} ( \mathbb{E} [{\bf Q}_j] - z +  \mathbb{E}  |{\bf Q}_j - z|),
 \label{eq:lemma1_p2}
\end{eqnarray}
where $\mathbb{E}_{\mathcal{A}_i}$ denotes the expectation over randomly selected $k_i$ storage nodes in $\mathcal{A}_i \subseteq \mathcal{S}$ according to probabilities $\pi_{i,1},\ldots, \pi_{i,m}$. From Cauchy-Schwarz inequality, we have
\begin{eqnarray}
\mathbb{E}  |{\bf Q}_j - z| \le  \sqrt{ (\mathbb{E} [{\bf Z}_j] - z)^2 + {\rm Var}[{\bf Q}_j] } \label{eq:lemma1_p3} .
\end{eqnarray}
Combining (\ref{eq:lemma1_p2}) and (\ref{eq:lemma1_p3}), we obtain the desired result by taking a minimization over $z\in\mathbb{R}$.

Finally, it is easy to verify that the bound is tight for the same binary distribution constructed in \cite{OS:12}, i.e., ${\bf Q}_j=z \pm \sqrt{ (\mathbb{E} [{\bf Q}_j] - z)^2 + {\rm Var}[{\bf Q}_j] }$ with probabilities:
\begin{eqnarray}
P_{+}= \frac{1}{2} + \frac{1}{2} \cdot \frac{\mathbb{E} [{\bf Q}_j] - z}{\sqrt{ (\mathbb{E} [{\bf Q}_j] - z)^2 + {\rm Var}[{\bf Q}_j] }} , \label{eq:lemma1_p4}
\end{eqnarray}
and $P_{-}=1-P_{+}$, which satisfy the mean and variance conditions. Therefore, the upper bound in (\ref{eq:lemma1}) is tight for this binary distribution.
$\square$
\vspace{-.15in}


\subsection{Derivation of Problem JLCM}
\noindent {\em Proof:} Plugging the results from Lemma~\ref{th:lemma_1} and Lemma~\ref{th:lemma_2} into (\ref{eq:JLRM-SC}) and applying the same $z$ to all $\bar{T}_i$ (which relax the problem and maintains inequality (\ref{eq:lemma1})), we obtain the desired objective function Problem JLCM. In the derivation, we used the fact that $\Lambda_j = \sum_i \lambda_i\pi_{i,j} \ \forall j$ from (\ref{eq:lambda}). Notice that the first summation is changed from $j\in\mathbb{S}_i$ in Lemma~\ref{th:lemma_1} to $j=\{1,\ldots,m\}$ because we should always assign $\pi_{i,j}=0$ to storage node $j$ that does not host any chunks of file $i$, i.e., for all $j\notin \mathcal{S}_i$. $\square$

\vspace{-.15in}

\subsection{Proof of Lemma \ref{th:lemma_3}}
\noindent {\em Proof:} Plugging $\rho_j = \Lambda_j/\mu_j$ into (\ref{eq:c1}) and (\ref{eq:c2}), it is easy to verify that $X_j$ and $Y_j$ are both convex in $\Lambda_j\in [0,\mu_j]$, i.e.,
\begin{eqnarray}
& & \frac{\partial ^2 X_j}{d \Lambda_j^2} = \frac{\mu_j^2 \Gamma_j^2}{\left( \mu_j  - \Lambda_j \right)^3} >0, \nonumber \\
& & \frac{\partial ^2 Y_j}{d  \Lambda_j^2} =\frac{2\mu_j^2\hat{\Gamma}_j^3}{3\left( \mu_j  - \Lambda_j \right)^3}+ \frac{\mu_j^4\Gamma_j^4(2\mu_j + 4\Lambda_j)} {(\mu_j - \Lambda_j)^4 } >0.  \nonumber
\end{eqnarray}
Similarly, we can verify that $G=\left[X_j + \sqrt{X_j^2 + Y_j} \right]$ is convex in $X_j$ and $Y_j$ by showing that its Hessian matrix is positive semi-definite, i.e.,
\begin{eqnarray}
\nabla^2 G = \frac{ 1  }{2\left(X_j^2+Y_j\right)^{\frac{3}{2}}} \cdot \left[\begin{array}{cc}
X_j^2 & X_j\\
 X_j & 1
\end{array} \right] \succeq {\bf 0}. \nonumber
\end{eqnarray}
Next, since $G$ is increasing in $X_j,Y_j$, their composition is convex and increasing in $\Lambda_j$ \cite{Boyd:05}. It further implies that $F=\Lambda_j G/2$ is also convex. This completes the proof. $\square$

\vspace{-.15in}


\subsection{Proof of Theorem \ref{th:thm_2}}
\noindent {\em Proof:} To simplify notations, we first introduce 2 auxiliary functions:
\begin{eqnarray}
& & g= \sum_{j=1}^m \frac{\Lambda_j}{2} \left[ X_j + \sqrt{X_j^2 + Y_j} \right], \\
& & h= \theta \sum_{i=1}^r \sum_{j=1}^m  \left[ V_j {\bf 1}_{\left( \pi_{i,j}^{(t)}>0\right) } + \frac{V_j(\pi_{i,j}- \pi_{i,j}^{(t)}) }{  (\pi_{\i,j}^{(t)} +1/\beta) \log\beta } \right]. \label{eq:111}
\end{eqnarray}
Therefore Problem (\ref{eq:cc0}) is equivalent to $\min_{\mathbf{\pi}} (g+h)$ over $\mathbf{\pi}=(\pi_{i,j}^{(t)} \ \forall i,j)$. For any $\beta>0$, due to the the concavity of logarithmic functions we have $\log(\beta y +1) - \log (\beta x +1) \le  \beta(y-x)/(\beta x +1)$ for any non negative $x,y$. Choosing $x=\pi_{i,j}^{(t)}$ and $y=\pi_{i,j}^{(t+1)}$ and multiplying a constant $V_j/\log\beta$ on both sides of the inequality, we have
\begin{eqnarray}
& \frac{V_j(\pi_{i,j}^{(t+1)}- \pi_{i,j}^{(t)}) }{  (\pi_{\i,j}^{(t)} +\frac{1}{\beta}) \log\beta } \ge V_j \frac{\log(\beta\pi_{i,j}^{(t+1)}+1 )}{\log\beta}  - V_j \frac{\log(\beta\pi_{i,j}^{(t)}+1 )}{\log\beta }.  & \label{eq:333}
\end{eqnarray}
Therefore we construct a new auxiliary function
\begin{eqnarray}
& & \hat{h} = \theta \sum_{i=1}^r \sum_{j=1}^m V_j \frac{\log(\beta\pi_{i,j}+1 )}{\log\beta } . \label{eq:111}
\end{eqnarray}
Since $\pi_{i,j}^{(t+1)}$ minimizes Problem (\ref{eq:cc0}), we have
\begin{eqnarray}
g(\pi^{(t+1)}) + h(\pi^{(t+1)}) \le g(\pi^{(t)}) + h(\pi^{(t)}). \label{eq:222}
\end{eqnarray}
Next we consider a new objective function $[g+\hat{h}]$ and show that it generates a descent sequence, i.e.,
\begin{eqnarray}
& & [g+\hat{h}](\pi^{(t+1)})  - [g+\hat{h}](\pi^{(t)}) \nonumber \\
&  &  \ \ \ \ \ \le h(\pi^{(t)}) - h(\pi^{(t+1)})  +\hat{h}(\pi^{(t+1)}) -\hat{h}(\pi^{(t)}) \nonumber \\
& & \ \ \ \ \ = \sum_{i=1}^r\sum_{j=1}^m \frac{V_j(\pi_{i,j}- \pi_{i,j}^{(t)}) }{  (\pi_{\i,j}^{(t)} +1/\beta) \log\beta } + \hat{h}(\pi^{(t+1)}) -\hat{h}(\pi^{(t)}) \nonumber \\
& & \ \ \ \ \ \le 0,
\end{eqnarray}
where the first step uses (\ref{eq:222}) and the last step follows from (\ref{eq:333}). Therefore, Algorithm JLCM generates a descent sequence, $\pi_{i,j}^{(t)}$ for $t=0,1,\ldots$, for objective function $[g+\hat{h}]$. Notice that for any $\pi_{i,j}\in[0,1]$, we have
\begin{eqnarray}
\lim_{\beta\rightarrow \infty} \hat{h}(\pi) =  \sum_{i=1}^r \sum_{j=1}^m V_j {\bf 1}_{\left( \pi_{i,j}>0\right) },\label{eq:555}
\end{eqnarray}
which is exactly the cost function in Problem JLCM. The converging point of the descent sequence is also a local optimal point of Problem JLCM as $\beta \rightarrow \infty$. $\square$

\end{document}

%% file: 1_intro.tex

\section{Introduction}

\label{sec:intro}

\subsection{Motivation}

Consumers are engaged in more social networking and E-commerce activities these days and are increasingly storing their documents and media in the online storage. Businesses are relying on Big Data analytics for business intelligence and are migrating their traditional IT infrastructure to the cloud.  These trends cause the online data storage demand to rise faster than Moore's Law \cite{dummy}. The increased storage demands have led companies to launch cloud storage services like Amazon's S3 \cite{AmazonS3} and personal cloud storage services like Amazon's Cloud drive, Apple's iCloud, DropBox, Google Drive, Microsoft's SkyDrive, and AT\&T Locker. Storing redundant information on distributed servers can increase reliability for storage systems, since users can retrieve duplicated pieces in case of disk, node, or site failures.

Erasure coding has been widely studied for distributed storage systems \cite[and references therein]{Dimakis} and used by companies like Facebook \cite{Sathiamoorthy13} and Google \cite{Fikes10} since it provides space-optimal data redundancy to protect against data loss. There is, however, a critical factor that affects the service quality that the user experiences, which is the delay in accessing the stored file. In distributed storage, the bandwidth between different nodes is frequently limited and so is the bandwidth from a user to different storage nodes, which can cause a significant delay in data access and perceived as poor quality of service. In this paper, we consider the problem of jointly minimizing both service delay and storage cost for the end users.

\begin{figure}[!thbp]
\begin{center}
\scalebox{0.59}{\includegraphics[draft=false]{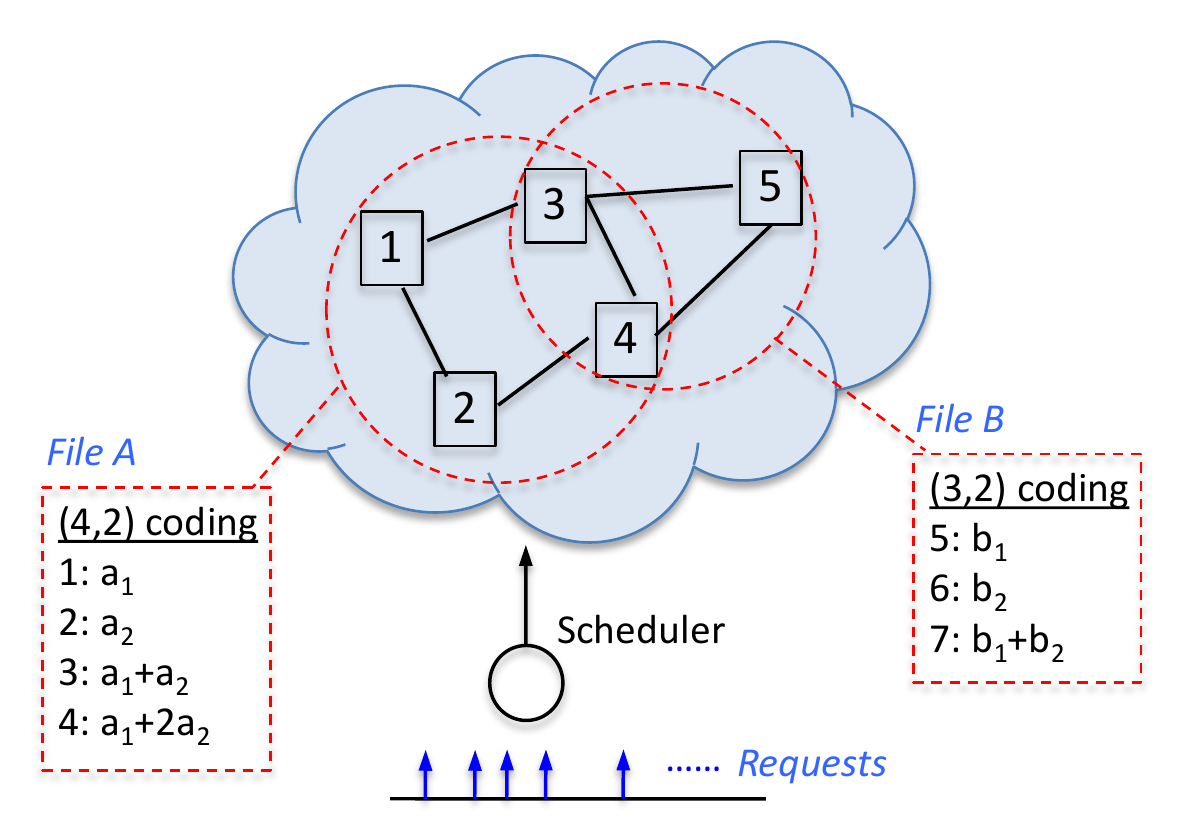}}
\caption{An erasure-coded storage of 2 files, which partitioned into 2 blocks and encoded using $(4,2)$ and $(3,2)$ MDS codes, respectively. Resulting file chunks are spread over 7 storage nodes. Any file request must be processed by 2 distinct nodes that have the desired chunks. Nodes $3,4$ are shared and can process request for both files.}
\label{fig:sysmodel}
\end{center}
\vspace{-.3in}
\end{figure}

While a latency-cost tradeoff is demonstrated for the special case of a single file, or homogeneous files with exactly the same properties(file size, type, coding parameters, etc.) \cite{ISIT:12,MG1:12,Joshi:13,MDS-Queue}, much less is known about the latency performance of multiple heterogeneous files that are coded with different parameters and share common storage servers. The main goal of this paper can be illustrated by an abstracted example shown in Fig.~\ref{fig:sysmodel}. We consider two files, each partitioned into $k=2$ blocks of equal size and encoded using maximum distance separable (MDS) codes. Under an $(n,k)$ MDS code, a file is encoded and stored in $n$ storage nodes such that the chunks stored in any $k$ of these $n$ nodes suffice to recover the entire file. There is a centralized scheduler that buffers and schedules all incoming requests. For instance, a request to retrieve file $A$ can be completed after it is successfully processed by 2 distinct nodes chosen from $\{1,2,3,4\}$ where desired chunks of $A$ are available. Due to shared storage nodes and joint request scheduling, delay performances of the files are highly correlated and are collectively determined by control variables of both files over three dimensions: (i) the scheduling policy that decides what request in the buffer to process when a node becomes available, (ii) the placement of file chunks over distributed storage nodes, and (iii) erasure coding parameters that decides how many chunks are created. A joint optimization over these three dimensions is very challenging because the latency performance of different files are tightly entangled. While increasing erasure code length of file B allows it to be placed on more storage nodes, potentially leading to smaller latency (because of improved load-balancing) at the price of higher storage cost, it inevitably affects service latency of file A due to resulting contention and interference on more shared nodes. In this paper, we present a quantification of service latency for erasure-coded storage with multiple heterogeneous files and propose an efficient solution to the joint optimization of both latency and storage cost.

\vspace{-.2in}
\subsection{Related Work}

The effect of coding on content retrieval latency in data-center storage system is drawing more and more significant attention these days, as Google and Amazon have published that every 500 ms extra delay means a 1.2\% user loss \cite{Google}. However, to our best knowledge quantifying the exact service delay in an erasure-coded storage system is an open problem, prior works focusing on asymptotic queuing delay behaviors \cite{Bramson:10,Lu:10} are not applicable because redundancy factor in practical data centers typically remain small due to storage cost concerns. Due to the lack of analytical latency models for erasure-coded storage, most of the literature is focused on reliable distributed storage system design, and latency is only presented as a performance metric when evaluating the proposed erasure coding scheme, e.g., \cite{DPR04,AJX05,HS07,WXHH06,J06}, which demonstrate latency improvement due to erasure coding in different system implementations. Related design can also be found in data access scheduling \cite{SH07,A98,TI10,RL05}, access collision avoidance \cite{KL98,ZA02}, and encoding/decoding time optimization \cite{SX,WK} and there are also some work using the LT erasure codes to adjust the system to meet user requirements such as availability, integrity and confidentiality \cite{AG14}. Restricting to the special case of a single file or homogeneous files, service delay bounds of erasure-coded storage have been recently studied in \cite{ISIT:12,MG1:12,Joshi:13,MDS-Queue}.

{\bf Queuing-theoretic analysis.}  For a single file or multiple but homogeneous files, under an assumption of exponential service time distribution, the authors in \cite{SV14} proved an asymptotic result for symmetric large-scale systems which can be applied to provide a computable approximation for expected latency, however, under a assumption that chunk placement is fixed and so is coding policy for all requests, which is not the case in reality. Also, the authors in \cite{ISIT:12,MG1:12} proposed a {\em block-one-scheduling} policy that only allows the request at the head of the buffer to move forward. An upper bound on the average latency of the storage system is provided through queuing-theoretic analysis for MDS codes with $k = 2$.  Later, the approach is extended in \cite{MDS-Queue} to general $(n,k)$ erasure codes, yet for a single file or homogeneous files. A family of {\em MDS-Reservation($t$)} scheduling policies that block all except the first $t$ of file requests are proposed and lead to numerical upper bounds on the average latency. It is shown that as $t$ increases, the bound becomes tighter while the number of states concerned in the queueing-theoretic analysis grows exponentially.

{\bf Fork-join queue analysis.} A queuing model closely related to erasure-coded storage is the fork-join queue \cite{Makowski:89} which has been extensively studied in the literature. Recently, in \cite{JK13}, the authors proposed a heuristic transmission scheme using this Fork-join queuing model where a file request is forked to all $n$ storage nodes that host the file chunks, and it exits the system when any $k$ chunks are processed to dynamically tuning coding parameters to improve latency performance. In \cite{LK13} the authors proposed a self-adaptive strategy which can dynamically adjusting chunk size and number of redundancy requests according to dynamic workload status in erasure-coded storage systems to minimize queuing delay in fork-join queues. Also the authors in \cite{Joshi:13} used this $(n,k)$ fork-join queue to model the latency performance of erasure-coded storage, a closed-form upper bound of service latency is derived for the case of a single file or homogeneous files and exponentially-distributed service time. However, the approach cannot be applied to a multiple-heterogeneous file storage where each file has a separate folk-join queue and the queues of different files are highly dependent due to shared storage nodes and joint request scheduling. In another work \cite{CS14}, the authors applied this fork-join queue  to optimize threads allocation to each request, which is similar to our weighted queue model, however, both proposed greedy/shared scheme would waste system resources because in fork-join queue there will always be some threads have unfinished downloads due to redundant assignment. In addition, in \cite{KT14}, the authors proposed a distributed storage system which analyzed through the Fork-join queue framework with heterogeneous jobs, and provide lower and upper bounds on the average latency for jobs of different classes under various scheduling policies, such as First-Come-First-Serve, preemptive and non-preemptive priority scheduling policies, based on the analysis of mean and second moment of waiting time. However, under a folk-join queue, each file request must be served by all $n$ nodes or a set of pre-specified nodes. It falls short to address dynamic load-balancing of multiple heterogeneous files.

\vspace{-.2in}

\subsection{Our Contributions}

This paper aims to propose a systematic framework that  {\em (i) quantifies the outer bound on the service latency of arbitrary erasure codes and for any number of files} in distributed data center storage with general service time distributions, and {\em (ii) enables a novel solution to a joint minimization of latency and storage cost} by optimizing the system over three dimensions: erasure coding, chunk placement, and scheduling policy.

The outer bound on the service latency is found using four steps, (i) We present a novel {\em probabilistic scheduling} policy, which dispatches each file request to $k$ distinct storage nodes who then manages their own local queues independently. A file request exits the system when all the $k$ chunk requests are processed. We show that probabilistic scheduling provides an upper bound on average latency of erasure-coded storage for arbitrary erasure codes, any number of files, and general services time distributions. (ii) Since the latency for probabilistic scheduling for all probabilities over ${n \choose k}$ subsets is hard to evaluate, we show that the probabilistic scheduling is equivalent to accessing each of the $n$ storage node with certain probability. If there is a strategy that accesses each storage node with certain probability, there exist a probabilistic scheduling strategy over all ${n \choose k}$ subsets. (iii) The policy that selects each storage node with certain probability generates memoryless requests at each of the node and thus the delay at each storage node can be characterized by the latency of M/G/1 queue. (iv) Knowing the exact delay from each storage node, we find a tight bound on the delay of the file by extending ordered statistic analysis in \cite{OS:12}. Not only does our result supersede previous latency analysis \cite{ISIT:12,MG1:12,Joshi:13,MDS-Queue} by incorporating multiple heterogeneous files and arbitrary service time distribution, it is also shown to be tighter for a wide range of workloads even in the single-file or homogeneous files case.

Multiple extensions to the outer bound on the service latency are considered. The first is the case when multiple chunks can be placed on the same node. As a result, multiple chunk requests corresponding to the
same file request can be submitted to the same queue, which
processes the requests sequentially and results in dependent
chunk service times. The second is the case when the file can be retrieved from more than $k$ nodes. In this case, smaller amount of data can be obtained from more nodes. Obtaining data from more nodes has an effect of considering worst ordered statistics  having an effect on increasing latency, while the smaller file size from each of the node helping more parallelization, and thus decreasing latency. The optimal value of the number of disks to access can then be optimized. 

The main application of our latency analysis is a joint optimization of latency and storage cost for multiple-heterogeneous file storage over three dimensions: erasure coding, chunk placement, and scheduling policy. To the best of our knowledge, this is the first paper to explore all these three design degrees of freedoms and to optimize an aggregate latency-plus-cost objective for all end users in an erasure-coded storage. Solving such a joint optimization is known to be hard due to the integer property of storage cost, as well as the coupling of control variables. While the length of erasure code determines not only storage cost but also the number of file chunks to be created and placed, the placement of file chunks over storage nodes further dictates the possible options of scheduling future file requests. To deal with these challenges, we propose an algorithm that constructs and computes a sequence of local, convex approximations of the latency-plus-cost minimization that is a mixed integer optimization. The sequence of approximations parametrized by $\beta>0$ can be efficiently computed using a standard projected gradient method and is shown to converge to the original problem as $\beta\rightarrow\infty$.



To validate our theoretical analysis and joint latency-plus-cost optimization, we provide a prototype of the proposed algorithm in {\em Tahoe} \cite{Tahoe}, which is an open-source, distributed filesystem based on the {\em zfec} erasure coding library for fault tolerance. A Tahoe storage system consisting of 12 storage nodes are deployed as virtual machines in an OpenStack-based data center environment distributed in New Jersey (NJ), Texas (TX), and California (CA).  Each site has four storage servers. One additional storage client was deployed in the NJ data center to issue storage requests. First, we validate our latency analysis via experiments with multiple-heterogeneous files and different request arrival rates on the testbed. Our measurement of real service time distribution falsifies the exponential assumption in \cite{ISIT:12,MG1:12,MDS-Queue}. Our analysis outperforms the upper bound in \cite{Joshi:13} even in the single-file/homogeneous-file case. Second, we implement our algorithm for joint latency-plus-cost minimization and demonstrate significant improvement of both latency and cost over oblivious design approaches. Our entire design is validated in various scenarios on our testbed, including different files sizes and arrival rates. The percentage improvement increases as the file size increases because our algorithm reduces queuing delay which is more effective when file sizes are larger. Finally, we quantify the tradeoff between latency and storage cost. It is shown that the improved latency shows a diminished return as storage cost/redundancy increase, suggesting the importance of identifying a particular tradeoff point.



%% file: 3_model_v2.tex
\section{System Model}
\label{sec:sys}
\vspace{-.1in}

We consider a data center consisting of $m$ heterogeneous servers, denoted by $\mathcal{M}=\{1,2,\ldots,m\}$, called storage nodes. To distributively store a set of $r$ files, indexed by $i=1,\ldots,r$, we partition each file $i$ into $k_i$ fixed-size chunks\footnote{While we make the assumption of fixed chunk size here to simplify the problem formulation, all results in this paper can be easily extended to variable chunk sizes. Nevertheless, fixed chunk sizes are indeed used by many existing storage systems \cite{DPR04,AJX05,LC02}.} and then encode it using an $(n_i,k_i)$ MDS erasure code to generate $n_i$ distinct chunks of the same size for file $i$. The encoded chunks are assigned to and stored on $n_i$ distinct storage nodes, which leads to a {\em chunk placement subproblem}, i.e., to find a set $\mathcal{S}_i$ of storage nodes, satisfying $\mathcal{S}_i\subseteq\mathcal{M}$ and $n_i=|\mathcal{S}_i|$, to store file $i$. Therefore, each chunk is placed on a different node to provide high reliability in the event of node or network failures. While data locality and network delay have been one of the key issues studied in data center scheduling algorithms \cite{TI10,RL05,RCCK12}, the prior work does not apply to erasure-coded systems.

The use of $(n_i,k_i)$ MDS erasure code allows the file to be reconstructed from any subset of $k_i$-out-of-$n_i$ chunks, whereas it also introduces a redundancy factor of $n_i/k_i$. To model storage cost, we assume that each storage node $j\in\mathcal{M}$ charges a constant cost $V_j$ per chunk. Since $k_i$ is determined by file size and the choice of chunk size, we need to choose an appropriate $n_i$ which not only introduces sufficient redundancy for improving chunk availability, but also achieves a cost-effective solution. We refer to the problem of choosing $n_i$ to form a proper $(n_i,k_i)$ erasure code as an {\em erasure coding subproblem}.

For known erasure coding and chunk placement, we shall now describe a queueing model of the distributed storage system. We assume that the arrival of client requests for each file $i$ form an independent Poisson process with a known rate $\lambda_i$. We consider chunk service time ${\bf X}_j$ of node $j$ with {\em arbitrary distributions}, whose statistics can be obtained inferred from existing work on network delay \cite{AY11,WK} and file-size distribution \cite{D11,PT12}. Under MDS codes, each file $i$ can be retrieved from any $k_i$ distinct nodes that store the file chunks. We model this by treating each file request as a {\em batch} of $k_i$ chunk requests, so that a file request is served when all $k_i$ chunk requests in the batch are processed by distinct storage nodes. All requests are buffered in a common queue of infinite capacity.

\begin{figure}[!thbp]
\begin{center}
\scalebox{0.32}{\includegraphics[draft=false]{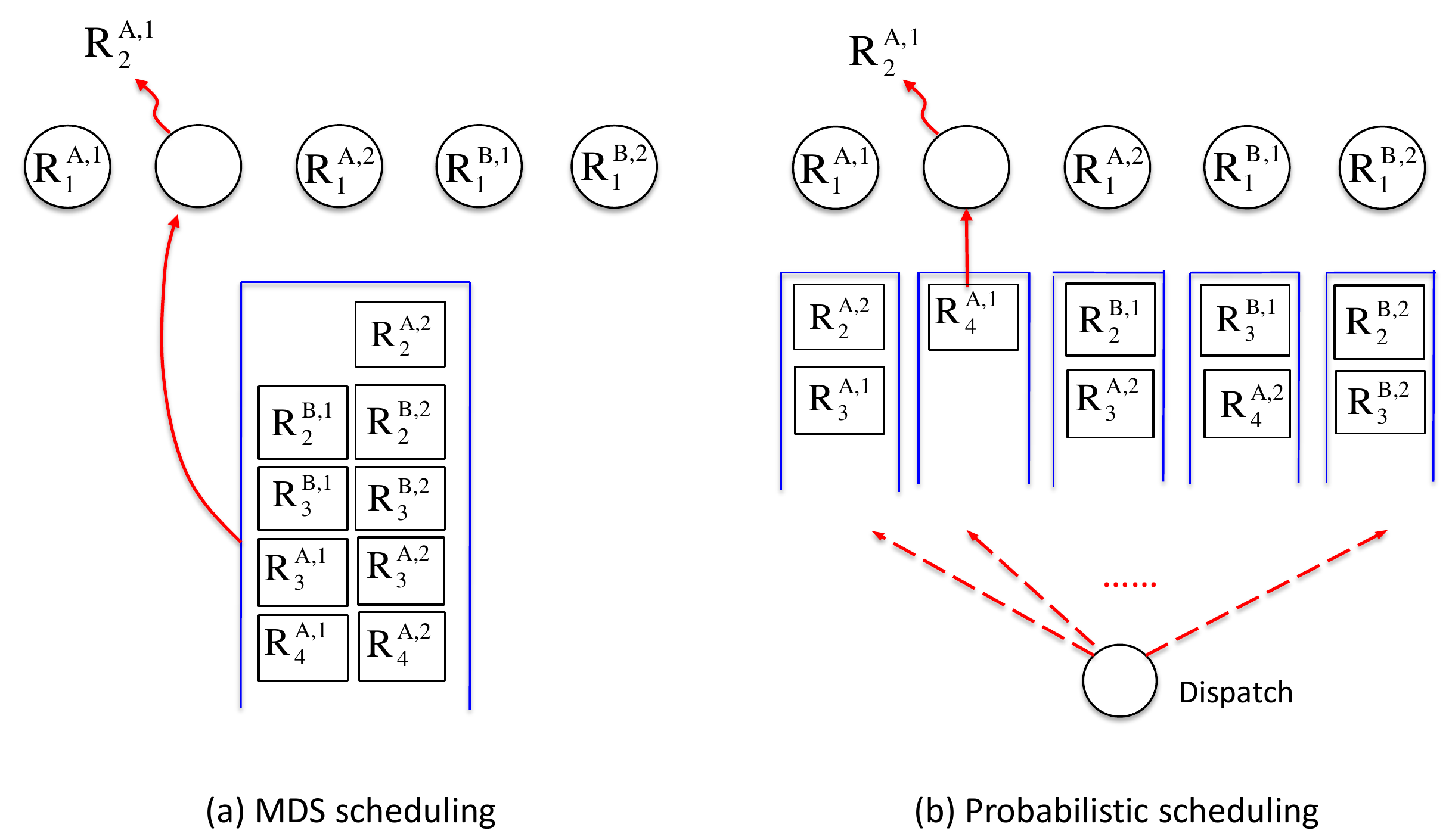}}
\caption{Functioning of (a) an optimal scheduling policy and (b) a probabilistic scheduling policy.}
\label{fig:sysmodel2}
\end{center}
\vspace{-.2in}
\end{figure}

Consider the 2-file storage example in Section~\ref{sec:intro}, where files $A$ and $B$ are encoded using $(4,2)$ and $(3,2)$ MDS codes, respectively, file $A$ will have chunks as $A_1$, $A_2$, $A_3$ and $A_4$, and file $B$ will have chunks $B_1$, $B_2$ and $B_3$. As depicted in Fig.\ref{fig:sysmodel2} (a), each file request comes in as a batch of $k_i=2$ chunk requests, e.g., $(R_1^{A,1},R_1^{A,2})$, $(R_2^{A,1},R_2^{A,2})$, and $(R_1^{B,1},R_1^{B,2})$, where $R_{i}^{A,j}$, denotes the $i$th request of file $A$, $j=1, 2$ denotes the first or second chunk request of this file request. Denote the five nodes (from left to right) as servers 1, 2, 3, 4, and 5, and we initial 4 file requests for file $A$ and 3 file requests for file $B$, i.e., requests for the different files have different arrival rates.  The two chunks of one file request can be any two different chunks from $A_1$, $A_2$, $A_3$ and $A_4$ for file $A$ and  $B_1$, $B_2$ and $B_3$ for file $B$. Due to chunk placement in the example, any 2 chunk requests in file A's batch must be processed by 2 distinct nodes from $\{1,2,3,4\}$, while 2 chunk requests in file B's batch must be served by 2 distinct nodes from $\{3,4,5\}$. Suppose that the system is now in a state depicted by Fig.\ref{fig:sysmodel2} (a), wherein the chunk requests $R_1^{A,1}$, $R_2^{A,1}$, $R_1^{A,2}$, $R_1^{B,1}$, and $R_2^{B,2}$ are served by the 5 storage nodes, and there are 9 more chunk requests buffered in the queue. Suppose that node 2 completes serving chunk request $R_2^{A,1}$ and is now free to server another request waiting in the queue. Since node 2 has already served a chunk request of batch $(R_2^{A,1},R_2^{A,2})$ and node 2 does not host any chunk for file B, it is not allowed to serve either $R_2^{A,2}$ or $R_2^{B,j},R_3^{B,j}$ where $j=1,2$ in the queue. One of the valid requests, $R_3^{A,j}$ and $R_4^{A,j}$, will be selected by an scheduling algorithm and assigned to node 2. We denote the scheduling policy that minimizes average expected latency in such a queuing model as {\em optimal scheduling}.

\begin{definition} {\em (Optimal scheduling)} An optimal scheduling policy (i) buffers all requests in a queue of infinite capacity; (ii) assigns at most 1 chunk request from a batch to each appropriate node, and (iii) schedules requests to minimize average latency if multiple choices are available.
\end{definition}

An exact analysis of optimal scheduling is extremely difficult. Even for given erasure codes and chunk placement, it is unclear what scheduling policy leads to minimum average latency of multiple heterogeneous files. For example, when a shared storage node becomes free, one could schedule either the earliest valid request in the queue or the request with scarcest availability, leading to different implications on average latency. A scheduling policy similar to \cite{ISIT:12,MG1:12} that blocks all but the first $t$ batches does not apply to multiple heterogeneous files because a Markov-chain representation of the resulting queue is required to have each state encapsulating not only the status of each batch in the queue, but also the exact assignment of chunk requests to storage nodes, since nodes are shared by multiple files and are no longer homogeneous. This leads to a Markov chain which has a huge state space and is hard to quantify analytically even for small $t$. On the other hand, the approach relying on $(n,k)$ fork-join queue in \cite{Joshi:13} also falls short because each file request must be forked to $n_i$ servers, inevitably causing conflict at shared servers.

%% file: 4_bound_v2.tex
\section{Upper bound: Probabilistic \\scheduling}
\label{sec:analysis}

This section presents a class of scheduling policies (and resulting latency analysis), which we call the {\rm probabilistic scheduling}, whose average latency upper bounds that of optimal scheduling.
\vspace{-.15in}

\subsection{Probabilistic scheduling}

Under $(n_i,k_i)$ MDS codes, each file $i$ can be retrieved by processing a batch of $k_i$ chunk requests at distinct nodes that store the file chunks. Recall that each encoded file $i$ is spread over $n_i$ nodes, denoted by a set $\mathcal{S}_i$. Upon the arrival of a file $i$ request, in probabilistic scheduling we randomly dispatch the batch of $k_i$ chunk requests to $k_i$-out-of-$n_i$ storage nodes in $\mathcal{S}_i$, denoted by a subset $\mathcal{A}_i\subseteq \mathcal{S}_i$ (satisfying  $|\mathcal{A}_i|=k_i$) with predetermined probabilities. Then, each storage node manages its local queue independently and continues processing requests in order. A file request is completed if all its chunk requests exit the system. An example of probabilistic scheduling is depicted in Fig.\ref{fig:sysmodel2} (b), wherein 5 chunk requests are currently served by the 5 storage nodes, and there are 9 more chunk requests that are randomly dispatched to and are buffered in 5 local queues according to chunk placement, e.g., requests $B_2,B_3$ are only distributed to nodes $\{3,4,5\}$. Suppose that node 2 completes serving chunk request $A_2$. The next request in the node's local queue will move forward.

\begin{definition} {\em (Probabilistic scheduling)} An Probabilistic scheduling policy (i) dispatches each batch of chunk requests to appropriate nodes with predetermined probabilities; (ii) each node buffers requests in a local queue and processes in order.
\end{definition}

It is easy to verify that such probabilistic scheduling ensures that at most 1 chunk request from a batch to each appropriate node. It provides an upper bound on average service latency for the optimal scheduling since rebalancing and scheduling of local queues are not permitted. Let $\mathbb{P}(\mathcal{A}_i)$ for all $\mathcal{A}_i\subseteq \mathcal{S}_i$ be the probability of selecting a set of nodes $\mathcal{A}_i$ to process the $|\mathcal{A}_i|=k_i$ distinct chunk requests\footnote{It is easy to see that $\mathbb{P}(\mathcal{A}_i)=0$ for all $\mathcal{A}_i\nsubseteq \mathcal{S}_i$ and $|\mathcal{A}_i|=k_i$ because such node selections do not recover $k_i$ distinct chunks and thus are inadequate for successful decode.}.

\begin{lemma} \label{th:lemma_bound}
For given erasure codes and chunk placement, average service latency of probabilistic scheduling with feasible probabilities $\{\mathbb{P}(\mathcal{A}_i): \ \forall i,\mathcal{A}_i\}$ upper bounds the latency of optimal scheduling.
\end{lemma}

Clearly, the tightest upper bound can be obtained by minimizing average latency of probabilistic scheduling over all feasible probabilities $\mathbb{P}(\mathcal{A}_i)$ $\forall \mathcal{A}_i\subseteq \mathcal{S}_i$ and $\forall i$, which involves $\sum_i (n_i$-choose-$k_i)$ decision variables. We refer to this optimization as a {\em scheduling subproblem}. While it appears prohibitive computationally, we will demonstrate next that the optimization can be transformed into an equivalent form, which only requires $\sum_i n_i$ variables. The key idea is to show that it is sufficient to consider the conditional probability (denoted by $\pi_{i,j}$) of selecting a node $j$, given that a batch of $k_i$ chunk requests of file $i$ are dispatched. It is easy to see that for given $\mathbb{P}(\mathcal{A}_i)$, we can derive $\pi_{i,j}$ by
\begin{eqnarray}
\pi_{i,j} = \sum_{\mathcal{A}_i:\mathcal{A}_i\subseteq \mathcal{S}_i} \mathbb{P}(\mathcal{A}_i) \cdot {\bf 1}_{\{ j\in\mathcal{A}_i\}}, \ \forall i \label{eq:pi}
\end{eqnarray}
where ${\bf 1}_{\{ j\in\mathcal{A}_i\}}$ is an indicator function which equals to 1 if node $j$ is selected by $\mathcal{A}_i$ and 0 otherwise.

\begin{theorem} \label{th:thm_prob}
A probabilistic scheduling policy with feasible probabilities $\{\mathbb{P}(\mathcal{A}_i): \ \forall i,\mathcal{A}_i\}$ exists if and only if there exists conditional probabilities $\{\pi_{i,j}\in [0,1], \forall i,j\}$ satisfying
\begin{eqnarray}
\sum_{j=1}^m \pi_{i,j} = k_i \ \forall i \ {\rm and} \ \pi_{i,j}=0 \ {\rm if} \ j\notin \mathcal{S}_i. \label{eq:thm_prob}
\end{eqnarray}
\end{theorem}

The proof of Theorem~\ref{th:thm_prob} relying on Farkas-Minkowski Theorem \cite{Angell:02} is detailed in the Appendix A. Intuitively, $\sum_{j=1}^m \pi_{i,j} = k_i$ holds because each batch of requests is dispatched to exact $k_i$ distinct nodes. Moreover, a node does not host file $i$ chunks should not be selected, meaning that $ \pi_{i,j}=0$ if $j\notin \mathcal{S}_i$. Using this result, it is sufficient to study probabilistic scheduling via conditional probabilities $\pi_{i,j}$, which greatly simplifies our analysis. In particular, it is easy to verify that under our model, the arrival of chunk requests at node $j$ form a Poisson Process with rate $\Lambda_j\sum_i \lambda_i\pi_{i,j}$, which is the superposition of $r$ Poisson processes each with rate $\lambda_i\pi_{i,j}$. The resulting queueing system under probabilistic scheduling is stable if all local queues are stable.

\begin{corollary} \label{th:stability}
The queuing system governed can be stabilized by a probabilistic scheduling policy under request arrival rates $\lambda_1,\lambda_2,\ldots,\lambda_r$ if there exists $\{\pi_{i,j}\in [0,1], \forall i,j\}$ satisfying (\ref{eq:pi_1}) and
\begin{eqnarray}
\Lambda_j = \sum_i \lambda_i\pi_{i,j} < \mu_j, \ \forall j. \label{eq:lambda}
\end{eqnarray}
\end{corollary}

\vspace{-.25in}

\subsection{Latency analysis and upper bound}

An exact analysis of the queuing latency of probabilistic scheduling is still hard because local queues at different storage nodes are dependent of each other as each batch of chunk requests are dispatched jointly. Let ${\bf Q}_j$ be the (random) waiting time a chunk request spends in the queue of node $j$. The expected latency of a file $i$ request is determined by the maximum latency that $k_i$ chunk requests experience on distinct servers,  $\mathcal{A}_i\subseteq \mathcal{S}_i$, which are randomly scheduled with predetermined probabilities, i.e.,
\begin{eqnarray}
\D \bar{T}_i= \mathbb{E}\left[\mathbb{E}_{\mathcal{A}_i} \left(\max_{ j\in \mathcal{A}_i} \{ {\bf Q}_j \} \right) \right], \label{eq:T_bar}
\end{eqnarray}
where the first expectation is taken over system queuing dynamics and the second expectation is taken over random dispatch decisions $\mathcal{A}_i$.

If the server scheduling decision $\mathcal{A}_i$ were deterministic, a tight upper bound on the expected value of the highest order statistic can be computed from marginal mean and variance of these random variables \cite{OS:12}, namely $\mathbb{E}[{\bf Q}_j]$ and Var$[{\bf Q}_j]$. Relying on Theorem~\ref{th:thm_prob}, we first extend this bound to the case of randomly selected servers with respect to conditional probabilities $\{\pi_{i,j}\in [0,1], \forall i,j\}$ to quantify the latency of probabilistic scheduling.

\begin{lemma} \label{th:lemma_1}
The expected latency $\bar{T}_i$ of file $i$ under probabilistic scheduling is upper bounded by
\begin{eqnarray}
\D & \bar{T}_i & \leq \min_{z\in \mathbb{R}} \left\{ z+\sum_{j\in \mathcal{S}_i} \frac{\pi_{i,j}}{2}  \left(\mathbb{E}[{\bf Q}_j] -z \right) \right.  \label{eq:lemma1} \nonumber\\
& & \left.+ \sum_{j\in \mathcal{S}_i} \frac{\pi_{i,j}}{2} \left[ \sqrt{(\mathbb{E}[{\bf Q}_j]-z)^2+{\rm Var}[{\bf Q}_j]}\right] \right\}.
\end{eqnarray}
The bound is tight in the sense that there exists a distribution of ${\bf Q}_j$ such that (\ref{eq:lemma1}) is satisfied with exact equality.
\end{lemma}

Next, we realize that the arrival of chunk requests at node $j$ form a Poisson Process with superpositioned rate $\Lambda_j=\sum_i \lambda_i\pi_{i,j}$. The marginal mean and variance of waiting time ${\bf Q}_j$ can be derived by analyzing them as separate M/G/1 queues. We denote ${\bf X}_j$ as the service time per chunk at node $j$, which has an arbitrary distribution satisfying finite mean $\mathbb{E}[{\bf X}_j]=1/\mu_j$, variance $\mathbb{E}[{\bf X}^2]-\mathbb{E}[{\bf X}]^2=\sigma^2_j$, second moment $\mathbb{E}[{\bf X}^2]=\Gamma_j^2$, and third moment $\mathbb{E}[{\bf X}^3]=\hat{\Gamma}^3_j$. These statistics can be readily inferred from existing work on network delay \cite{AY11,WK} and file-size distribution \cite{D11,PT12}.

\begin{lemma}\label{th:lemma_2}
Using Pollaczek-Khinchin transform \cite{MG1:12}, expected delay and variance for total queueing and network delay are given by
\begin{eqnarray}
\mathbb{E}[  {\bf Q}_j] =  \frac{1}{\mu_j} + \frac{ \Lambda_j \Gamma_j^2 }{2(1- \rho_j)}, \label{eq:lemma2_1}
\end{eqnarray}
\vspace{-0.2in}
\begin{eqnarray}
{\rm Var}[ {\bf Q}_j] =\sigma_j^2+\frac{ \Lambda_j \hat{\Gamma}_j^3}{3(1-\rho_j)}+\frac{\Lambda_j^2\Gamma_j^4}{4(1- \rho_j)^2} , \label{eq:lemma2_2}
\end{eqnarray}
where $\rho_j=\Lambda_j / \mu_j$ is the request intensity at node $j$.
\end{lemma}

Combining Lemma~\ref{th:lemma_1} and Lemma~\ref{th:lemma_2}, a tight upper bound on expected latency of file $i$ under probabilistic scheduling can be obtained by solving a single-variable minimization problem over real $z\in \mathbb{R}$ for given erasure codes $n_i$, chunk placement $\mathcal{S}_i$, and scheduling probabilities $\pi_{ij}$.

\vspace{0.05in}
\noindent{\em Remark 1:} Consider the homogeneous case studied in previous work \cite{ISIT:12,MG1:12,Joshi:13,CS14} where all nodes have the same service time distribution and where files have the same chunk placement (i.e., $|\mathcal{S}_i|=n_i=m$ $\forall i$ ). It is easy to show that due to symmetry, the optimal scheduling probabilities $\pi_{i,j}$ minimizing total system latency is $\pi_{i,j}=k_i/m$ for all $i,j$. Therefore, each node $j$ receives an equal request arrival rate $\Lambda_j$, resulting in equal mean and variance of waiting time ${\bf Q}_j$. Using the convexity of our bound with respect to $z$, the latency upper bound in \eqref{eq:lemma1} can be derived in closed form:
\begin{eqnarray}
\bar{T}_i\le \mathbb{E}[  {\bf Q}_j] + \sqrt{k_i-1} \cdot {\rm Var}[ {\bf Q}_j]
\end{eqnarray}
where $\mathbb{E}[  {\bf Q}_j]$ and ${\rm Var}[ {\bf Q}_j]$ are mean and variance of waiting time ${\bf Q}_j$ given by (\ref{eq:lemma2_1}) and (\ref{eq:lemma2_2}).

\vspace{-.15in}

\subsection{Extensions of the Latency Upper Bound}

In the above upper bound, we assumed that each file $i$ uses $(n_i,k_i)$ MDS code, places exactly one chunk on each selected node, and is retrieved from $k_i$ out of $n_i$ nodes on which the file is placed. In practice, more complicated storage schemes can be designed to offer a higher degree of elasticity by (i) placing multiple chunks on selected nodes or (ii) accessing the file from more than $k_i$ nodes in parallel. In this subsection, we further extend our latency upper bound to address these cases.

\noindent {\bf Placing multiple chunks on each node.} This case arises when a group of storage nodes share a single bottleneck (e.g., outgoing bandwidth at a regional datacenter) and must be modeled by a single queue, or in small clusters the number storage node is less than that of created file chunks (i.e., $n_i>m$). As a result, multiple chunk requests corresponding to the same file request can be submitted to the same queue, which processes the requests sequentially and results in dependent chunk service times.

To extend our latency bound, we assume that each node can host up to $c$ chunks. Thus, our probabilistic scheduling policy dispatches $x$ chunk requests of file $i$ to node $j$ with pre-determined probability $\hat{\pi}_{i,j}^{x}$ for $x=1,\ldots,c$. Since $\sum_x x\hat{\pi}_{i,j}^{x}$ represents the average number of chunks retrieved from node $j$, we must have $\sum_j\sum_x x\hat{\pi}_{i,j}^{x}=k_i$ to guarantee access to enough chunks for successful file retrieval. Further, these $x$ chunk requests join the service queue at the same time and have dependent service times, given by ${\bf Q}_j, {\bf Q}_j+{\bf X}_j^{1}, {\bf Q}_j+{\bf X}_j^{1}+{\bf X}_j^{2},\dots$ where ${\bf Q}_j$ is the waiting time of a single chunk request as before, and ${\bf X}_j^{1},{\bf X}_j^{2},\dots$ are {\em i.i.d.} chunk service times of node $j$, with mean $1/\mu_j$, variance $\sigma_j^2$, and third moment $\hat{\Gamma}_j^3$. Under this model, the latency of each file $i$ can be characterized by

\begin{lemma} \label{th:lemma_bound1}
The expected latency $\bar{T}_i$ of file $i$ is upper bounded by
\begin{eqnarray}
\D & \bar{T}_i & \leq \min_{z\in \mathbb{R}} \left\{ z+\sum_{j\in \mathcal{S}_i} \sum_{x=1}^c \frac{\hat{\pi}_{i,j}^{x}}{2}  \left(\mathbb{E}[\hat{\bf Q}_{ij}^{x}] -z \right) \right.   \nonumber\\
& & \left.+ \sum_{j\in \mathcal{S}_i}\sum_{x=1}^c \frac{\hat{\pi}_{i,j}^{x}}{2} \left[ \sqrt{(\mathbb{E}[\hat{\bf Q}_{ij}^{x}]-z)^2+{\rm Var}[\hat{\bf Q}_{ij}^{x}]}\right] \right\},
\end{eqnarray}
where $\hat{\bf Q}_{ij}^{x}$ is the waiting time for all $x$ chunk request of file $i$ submitted together to the queue of node $j$, with moments given as follows
\begin{eqnarray}
\mathbb{E}[  \hat{\bf Q}_{ij}^{x}] =  \frac{x}{\mu_j} + \frac{ \hat{\Lambda}_j \Gamma_j'^2 }{2(1- \rho_j)},
\end{eqnarray}
\vspace{-0.2in}
\begin{eqnarray}
{\rm Var}[ \hat{\bf Q}_{ij}^{x}] = x\sigma_j^2+\frac{ \hat{\Lambda}_j \hat{\Gamma}_j'^3}{3(1-\rho_j)}+\frac{\hat{\Lambda}_j^2\Gamma_j'^4}{4(1- \rho_j)^2} ,
\end{eqnarray}
where $\hat{\lambda}_j=\sum_i\sum_{x=1}^c x\hat{\pi}_{i,j}^{x}$ is the total request arrival rate at node $j$.
\end{lemma}

The proof is very similar to that of Lemma~\ref{th:lemma_bound}, recognizing that for all $x$ chunk requests served by the same queue, only the latency of last request (denoted by $\hat{\bf Q}_j^{x}={\bf Q}_j+{\bf X}_j^{1}+\ldots+{\bf X}_j^{x-1})$ has to be considered in the order statistic analysis, since it strictly dominates the queuing latency of other $x-1$ requests. Using the {\em i.i.d.} property of service times and updating total request arrival rate $\hat{\lambda}_j=\sum_i\sum_x x\hat{\pi}_{i,j}^{x}$, the proof of Lemma~\ref{th:lemma_bound1} is straightforward.

\noindent {\bf Retrieving file from more than $k_i$ nodes.} Let $F_i$ be the size of file $i$. We now consider the scenario where files have different chunk sizes and where each file $i$ can be obtained from $d_i\ge k_i$ nodes, requiring only $F_i/d_i$ amount of data from each node. The scheme allows a higher degree of parallelism in file access. Since less content is requested from each node, it may lead to lower service latency at the cost of accessing more nodes and more complicated coding strategy.

We first note that file can be retrieved by obtaining $F_i/d_i$ amount of data from $d_i\ge k_i$ nodes with the same placement and the same $(n_i,k_i)$ MDS code. To see this, consider that the content at each node is subdivided into $B=\binom{d_i}{k_i}$ sub-chunks (We assume that each chunk can be perfectly divided and ignore the effect of non-perfect division). Let ${\cal L} = \{{\cal L}_1, \cdots, {\cal L}_{ B}\}$ be the list of all ${B}$ combinations of $d_i$ servers  such that each combination is of size $k_i$. In order to access the data, we get $m^{\text{th}}$ sub-chunks from all the servers in ${\cal L}_m$ for all $m = 1, 2, \cdots {B}$. Thus, the total size of data retrieved is of size $F_i$, which is evenly accessed from all the $d_i$ nodes. In order to obtain the data, we have enough data to decode since $k_i$ sub-chunks are available for each $m$ and we assume a linear MDS code. We further assume that the if the service time for a chunk is proportional to its size, and thus if less content is requested from a server, the service time for that content is proportionally smaller.

\begin{lemma} \label{th:lemma_n}
The expected latency $\bar{T}_i$ of file $i$ is upper bounded by
\begin{eqnarray}
\D & \bar{T}_i & \leq \min_{z\in \mathbb{R}} \left\{ z+\sum_{j\in \mathcal{S}_i} \frac{\pi_{i,j}}{2}  \left(\mathbb{E}[\tilde{\bf Q}_{ij}] -z \right) \right.   \nonumber\\
& & \left.+ \sum_{j\in \mathcal{S}_i} \frac{\pi_{i,j}}{2} \left[ \sqrt{(\mathbb{E}[\tilde{\bf Q}_{ij}]-z)^2+{\rm Var}[\tilde{\bf Q}_{ij}]}\right] \right\},
\end{eqnarray}
where $\tilde{\bf Q}_{ij}$ is the (random) waiting time a chunk request for file $i$ spends in the queue of node $j$, with moments given as follows
\begin{eqnarray}
\mathbb{E}[  \tilde{\bf Q}_{ij}] =  \frac{k_i}{d_i\mu_j} + \frac{ \Lambda_j \Gamma_j'^2 }{2(1- \rho_j)},
\end{eqnarray}
\vspace{-0.2in}
\begin{eqnarray}
{\rm Var}[ \tilde{\bf Q}_{ij}] = \frac{k_i^2}{d_i^2}\sigma_j^2+\frac{ \Lambda_j \hat{\Gamma}_j'^3}{3(1-\rho_j)}+\frac{\Lambda_j^2\Gamma_j'^4}{4(1- \rho_j)^2} ,
\end{eqnarray}
where $\rho_j=\Lambda_j  \nu_j$ is the request intensity at node $j$, $\Lambda_j = \sum_i \lambda_i\pi_{i,j}, \ \forall j$, $0\le \pi_{i,j} \le 1$, $\sum_{j=1}^m \pi_{i,j} = d_i \ \forall i \ {\rm and} \ \pi_{i,j}=0 \ {\rm if} \ j\notin \mathcal{S}_i$,  and $k_i\le d_i \le n_i \ \forall i$. Further, $\nu_j = \sum_i \frac{\lambda_i\pi_{i,j}}{\sum_i \lambda_i\pi_{i,j}}\frac{k_i}{d_i\mu_j}$, $\Gamma_j'^2 = \sum_i \frac{\lambda_i\pi_{i,j}}{\sum_i \lambda_i\pi_{i,j}}\frac{k_i^2}{d_i^2} \Gamma_j^2$, and $\hat{\Gamma}_j'^3 = \sum_i \frac{\lambda_i\pi_{i,j}}{\sum_i \lambda_i\pi_{i,j}}\frac{k_i^3}{d_i^3} \hat{\Gamma}_j^3$.
\end{lemma}

The key step to prove Lemma~\ref{th:lemma_n} is to find the mean and variance of waiting time $\tilde{\bf Q}_{ij}$ for chunk requests on node $j$. Due to our assumption of proportional processing time, the service time of a file $i$ request is $k_i {\bf X}_j / d_i$ where ${\bf X}_j$ is the service time for a standard chunk size as before. Chunk requests submitted to node $j$ form a composite Poission process. Therefore, its service time follows the distribution of $k_i {\bf X}_j / d_i$ with normalized probabilities $\pi_{i,j}/(\sum_i \pi_{i,j})$ for $i=1,\dots, r$. The rest of the proof is straightforward by plugging this new service time distribution into the proof of Lemma~\ref{th:lemma_bound}.


%% file: 5_algo_v2.tex
\section{Application: Joint latency and cost optimization}
\label{sec:algo}

In this section, we address the following questions: what is the optimal tradeoff point between latency and storage cost for a erasure-coded system? While any optimization regarding exact latency is an open problem, the analytical upper bound using probabilistic scheduling enables us to formulate a novel optimization of joint latency and cost objectives. Its solution not only provides a theoretical bound on the performance of optimal scheduling, but also leads to implementable scheduling policies that can exploit such tradeoff in practical systems.

\vspace{-.15in}

\subsection{Formulating the Joint Optimization}

We showed that a probabilistic scheduling policy can be optimization over three sets of control variables: erasure coding $n_i$, chunk placement $\mathcal{S}_i$, and scheduling probabilities $\pi_{ij}$. However, a latency optimization without considering storage cost is impractical and leads to a trivial solution where every file ends up spreading over all nodes. To formulate a joint latency and cost optimization, we assume that storing a single chunk on node $j$ requires cost $V_j$, reflecting the fact that nodes may have heterogeneous quality of service and thus storage prices. Therefore, total storage cost is determined by both the level of redundancy (i.e., erasure code length $n_i$) and chunk placement  $\mathcal{S}_i$. Under this model, the cost of storing file $i$ is given by $C_i=\sum_{j\in \mathcal{S}_i} V_j$. In this paper, we only consider the storage cost of chunks while network cost would be an interesting future direction.

Let $\hat{\lambda}=\sum_i \lambda_i$ be the total arrival rate, so $\lambda_i/\hat{\lambda}$ is the fraction of file $i$ requests, and average latency of all files is given by $\sum_i (\lambda_i/\hat{\lambda})\bar{T}_i$. Our objective is to minimize an aggregate {\em latency-cost} objective, i.e.,
\begin{eqnarray}
& \D {\rm min} & \sum_{i=1}^r \frac{\lambda_i}{\hat{\lambda}}  \bar{T}_i + \theta \sum_{i=1}^r \sum_{j\in \mathcal{S}} V_j \label{eq:JLRM-SC} \\
& {\rm s.t.} & {\rm (\ref{eq:pi}), \ (\ref{eq:thm_prob}), \ (\ref{eq:lambda}), \ (\ref{eq:lemma1}), \ (\ref{eq:lemma2_1}), \ (\ref{eq:lemma2_2}). }  \label{eq:JLRM-SC2}   \nonumber \\
& {\rm var.} & n_i,  \ \pi_{i,j}, \ \mathcal{S}_i\in \mathcal{M}, \ \forall i,j. \nonumber
\end{eqnarray}
Here $\theta\in [0, \infty)$ is a tradeoff factor that determines the relative importance of latency and cost in the minimization problem. Varying from $\theta=0$ to $\theta \to \infty$, the optimization solution to (\ref{eq:JLRM-SC}) ranges from those minimizing latency to ones that achieve lowest cost.

The joint latency-cost optimization is carried out over three sets of variables: erasure code $n_i$, scheduling probabilities $\pi_{i,j}$, and chunk placement $\mathcal{S}_i$, subject to the constraints derived in Section~\ref{sec:analysis}. Varying $\theta$, the optimization problem allows service providers to exploit a latency-cost tradeoff and to determine the optimal operating point for different application demands. We plug into (\ref{eq:JLRM-SC}) the results in Section~\ref{sec:analysis} and obtain a Joint Latency-Cost Minimization (JLCM) with respect to probabilistic scheduling\footnote{The optimization is relaxed by applying the same axillary variable $z$ to all $\bar{T}_i$, which still satisfies the inequality (\ref{eq:lemma1}).}:

\noindent \hspace{0.2in} {\bf Problem JLCM:}
\vspace{-0.05in}
\begin{eqnarray}
& {\rm min} &  z + \sum_{j=1}^m \frac{\Lambda_j}{2\hat{\lambda}} \left[ X_j + \sqrt{X_j^2 + Y_j} \right] + \theta\sum_{i=1}^r \sum_{j\in \mathcal{S}_i} V_j \label{eq:c0} \\
& {s.t.} & X_j=   \frac{1}{\mu_j} + \frac{ \Lambda_j \Gamma_j^2 }{2(1- \rho_j)}-z, \ \forall j  \label{eq:c1}   \\
&  & Y_j=  \sigma_j^2+\frac{ \Lambda_j \hat{\Gamma}_j^3}{3(1-\rho_j)}+\frac{\Lambda_j^2\Gamma_j^4}{4(1- \rho_j)^2}, \ \forall j \label{eq:c2}  \\
& & \rho_j = \Lambda_j/\mu_j < 1; \ \Lambda_j=\sum_{i=1}^r \pi_{i,j}\lambda_i  \ \forall j \label{eq:c3} \\
&  & \sum_{j=1}^m \pi_{i,j} = k_i; \ \pi_{i,j}\in [0,1]; \ \pi_{i,j}=0 \ \forall j\notin \mathcal{S}_i \label{eq:c4}  \\
&  & |\mathcal{S}_i|=n_i \ {\rm and} \  \mathcal{S}_i\subseteq \mathcal{M}, \ \forall i \label{eq:c5}  \\
& {\rm var.} & z, \ n_i, \ \mathcal{S}_i, \ \pi_{i,j}, \ \forall i,j. \nonumber
\end{eqnarray}

\begin{figure*}
\hspace{0.1cm}
{\footnotesize \fbox{
\begin{minipage}{0.45\textwidth}
\begin{center}
\begin{tabbing}
xx\=xx\=xx\=xx\=xx\=xx\=\kill
{\bf Algorithm JLCM} : \\
\\
Choose sufficiently large $\beta>0$ \\
Initialize $t=0$ and feasible $(\pi_{i,j}^{(0)} \ \forall i,j)$\\
Compute current objective value $B^{(0)}$ \\
\textbf{while} $B^{(0)}-B^{(1)}>\epsilon$ \\
\> Approximate cost function using (\ref{eq:app}) and $(\pi_{i,j}^{(t)} \ \forall i,j)$ \\
\> {\bf Call} $projected\_gradient()$ to solve optimization (\ref{eq:cc0})\\
\> \> $(\pi_{i,j}^{(t+1)} \ \forall i,j) = \arg \min$ (\ref{eq:cc0}) \\
\> $z = \arg \min$ (\ref{eq:cc0}) \\
\> Compute new objective value $B^{(t+1)}$ \\
\> Update  $t=t+1$\\
{\bf end while} \\
Find chunk placement $\mathcal{S}_i$ and erasure code $n_i$ by (\ref{eq:lemma4}) \\
{\bf Output}: $(n_i, \mathcal{S}_i, \pi_{i,j}^{(t)})$ $\forall i,j$
\end{tabbing}
\label{fig:protocol}
\caption{Algorithm JLCM: Our proposed algorithm for solving Problem JLCM.}
\end{center}
\end{minipage}
}
\hspace{0.2cm}
\fbox{
\begin{minipage}{0.45\textwidth}
\begin{center}
\begin{tabbing}
xx\=xx\=xx\=xx\=xx\=xx\=\kill
{\bf Routine $projected\_gradient()$} : \\
\\
Choose proper stepsize $\delta_1,\delta_2, \delta_3,\ldots$ \\
Initialize $s=0$ and $\pi_{i,j}^{(s)} = \pi_{i,j}^{(t)}$\\
\textbf{while} $  \sum_{i,j} |\pi_{i,j}^{(s+1)}- \pi_{i,j}^{(s)}|    >\epsilon$ \\
\> Calculate gradient $\nabla$(\ref{eq:app}) with respect to $\pi_{i,j}^{(s)}$\\
\>  $\pi_{i,j}^{(s+1)}=\pi_{i,j}^{(s)}+\delta_s \cdot \nabla$(\ref{eq:app}) \\
\> {\bf Project}  $\pi_{i,j}^{(s+1)}$ onto feasibility set: \\
\> \> $\{\pi_{i,j}^{(s+1)}: \ \sum_j \pi_{i,j}^{s+1}=k_i, \  \pi_{i,j}^{s+1}\in[0,1], \ \forall i,j\}$ \\
\> Update  $s=s+1$\\
{\bf end while} \\
{\bf Output}: $(\pi_{i,j}^{(s)}, \ \forall i,j)$
\end{tabbing}
\label{fig:protocol0}
\caption{Projected Gradient Descent Routine, used in each iteration of Algorithm JLCM.}
\end{center}
\end{minipage}
}
}
\vspace{-.3in}
\end{figure*}

Problem JLCM is challenging due to two reasons. First, all optimization variables are highly coupled, making it hard to apply any greedy algorithm that iterative optimizes over different sets of variables. The number of nodes selected for chunk placement (i.e., $\mathcal{S}_i$) is determined by erasure code length $n_i$ in (\ref{eq:c5}), while changing chunk placement $\mathcal{S}_i$ affects the feasibility of probabilities $\pi_{i,j}$ due to (\ref{eq:c4}). Second, Problem JLCM is a mixed-integer optimization over $\mathcal{S}_i$ and $n_i$, and storage cost $C_i=\sum_{j\in \mathcal{S}_i} V_j$ depends on the integer variables. Such a mixed-integer optimization is known to be difficult in general

\vspace{-.15in}

\subsection{Constructing convex approximations}

In the next, we develop an algorithmic solution to Problem JLCM by iteratively constructing and solving a sequence of convex approximations. This section shows the derivation of such approximations for any given reference point, while the algorithm and its convergence will be presented later.

Our first step is to replace chunk placement $\mathcal{S}_i$ and erasure coding $n_i$ by indicator functions of $\pi_{i,j}$. It is easy to see that any nodes receiving a zero probability $\pi_{i,j}=0$ should be removed from $\mathcal{S}_i$, since any chunks placed on them do not help reducing latency.

\begin{lemma}\label{th:lemma_4}
The optimal chunk placement of Problem JLCM must satisfy $\mathcal{S}_i= \{j: \pi_{i,j}>0 \}$ $\forall i$, which implies
\begin{eqnarray}
\sum_{j\in \mathcal{S}_i} V_j  = \sum_{j=1}^m V_j {\bf 1}_{(\pi_{i,j}>0)}, \ \ n_i= \sum_{j=1}^m V_j {\bf 1}_{(\pi_{i,j}>0)}
\label{eq:lemma4}
\end{eqnarray}
\end{lemma}

Thus, Problem JLCM becomes to an optimization over only $(\pi_{i,j} \ \forall i,j)$, constrained by $\sum_{j=1}^m \pi_{i,j} = k_i \ {\rm and} \ \pi_{i,j}\in [0,1]$ in (\ref{eq:c4}), with respect to the following objective function:
\begin{eqnarray}
z+ \sum_{j=1}^m \frac{\Lambda_j}{2\bar{\lambda}} \left[ X_j + \sqrt{X_j^2 + Y_j} \right]+\theta \sum_{i=1}^r \sum_{j=1}^m V_j {\bf 1}_{(\pi_{i,j}>0)}. \label{eq:app1}
\end{eqnarray}
However, the indicator functions above that are neither continuous nor convex. To deal with them, we select a fixed reference point $(\pi_{i,j}^{(t)} \ \forall i,j)$ and leverage a linear approximation of (\ref{eq:app1}) with in a small neighbourhood of the reference point. For all $i,j$, we have
\begin{eqnarray}
 V_j {\bf 1}_{(\pi_{i,j}>0)}  \approx \left[ V_j {\bf 1}_{\left( \pi_{i,j}^{(t)}>0\right) } + \frac{V_j(\pi_{i,j}- \pi_{i,j}^{(t)}) }{  (\pi_{\i,j}^{(t)} +1/\beta) \log\beta } \right], \label{eq:app}
\end{eqnarray}
where $\beta>0$ is a sufficiently large constant relating to the approximation ratio. It is easy to see that the approximation approaches the real cost function within a small neighbourhood of $(\pi_{i,j}^{(t)} \ \forall i,j)$ as $\beta$ increases. More precisely, when $\pi_{i,j}^{(t)}=0$ the approximation reduces to $\pi_{i,j}(V_j\beta/\log\beta)$, whose gradient approaches infinity as $\beta\rightarrow \infty$, whereas the approximation converges to constant $V_j$ for any $\pi_{i,j}^{(t)}=0$ as $\beta\rightarrow \infty$.

It is easy to verify that the approximation is linear and differentiable. Therefore, we could iteratively construct and solve a sequence of approximated version of Problem JLCM. Next, we show that the rest of optimization objective in (\ref{eq:c0}) is convex in $\pi_{i,j}$ when all other variables are fixed.

\begin{lemma}\label{th:lemma_3}
The following function, in which $X_j$ and $Y_j$ are functions of $\Lambda_j$ defined by (\ref{eq:c1}) and (\ref{eq:c2}), is convex in $\Lambda_j$:
\begin{eqnarray}
 F(\Lambda_j) = \frac{\Lambda_j}{2\hat{\lambda}} \left[X_j + \sqrt{X_j^2 + Y_j} \right]. \label{eq:lemma3}
\end{eqnarray}
\end{lemma}

\vspace{-.2in}

\subsection{Algorithm JLCM and convergence analysis}

Leveraging the linear local approximation in (\ref{eq:app}) our idea to solve Problem JLCM is to start with an initial $(\pi_{i,j}^{(0)} \ \forall i,j)$, solve its optimal solution, and iteratively improve the approximation by replacing the reference point with an optimal solution computed from the previous step. Lemma~\ref{th:lemma_3} shows that such approximations of Problem JLCM are convex and can be solved by off-the-shelf optimization tools, e.g., Gradient Descent Method and Interior Point Method \cite{Boyd:05}.

The proposed algorithm is shown in Figure~\ref{fig:protocol}. For each iteration $t$, we solve an approximated version of Problem JLCM over $(\pi_{i,j}^{(0)} \ \forall i,j)$  with respect to a given reference point and a fixed parameter $z$. More precisely,  for $t=1,2,\ldots$ we solve
\begin{eqnarray}
& {\rm min} & \theta \sum_{i=1}^r \sum_{j=1}^m  \left[ V_j {\bf 1}_{\left( \pi_{i,j}^{(t)}>0\right) } + \frac{V_j(\pi_{i,j}- \pi_{i,j}^{(t)}) }{  (\pi_{\i,j}^{(t)} +1/\beta) \log\beta } \right] \nonumber \\
& & \ \ \ + z + \sum_{j=1}^m \frac{\Lambda_j}{2\hat{\lambda}} \left[ X_j + \sqrt{X_j^2 + Y_j} \right]   \label{eq:cc0}  \\
& {s.t.} & {\rm Constraints \ (\ref{eq:c1}), \ (\ref{eq:c2}), \ (\ref{eq:c3}) } \nonumber \\
& &  \sum_{j=1}^m \pi_{i,j} = k_i \ {\rm and} \ \pi_{i,j} \in [0,1] \nonumber  \\
& { var.} & \pi_{i,j} \ \forall i,j. \nonumber
\end{eqnarray}
Due to Lemma~\ref{th:lemma_3}, the above minimization problem with respect to a given reference point has a convex objective function and linear constraints. It is solved by a projected gradient descent routine in Figure~\ref{fig:protocol0}. Notice that the updated probabilities $(\pi_{i,j}^{(t)} \ \forall i,j)$ in each step are projected onto the feasibility set $\{\sum_j \pi_{i,j}=k_i, \  \pi_{i,j}\in[0,1], \ \forall i,j\}$ as required by Problem JLCM using a standard Euclidean projection. It is shown that such a projected gradient descent method solves the optimal solution of Problem (\ref{eq:cc0}). Next, for fixed probabilities $(\pi_{i,j}^{(t)} \ \forall i,j)$, we improve our analytical latency bound by minimizing it over $z\in \mathbb{R}$. The convergence of our proposed algorithm is proven in the following theorem.

\begin{theorem}\label{th:thm_2}
Algorithm JLCM generates a descent sequence of feasible points, $\pi_{i,j}^{(t)}$ for $t=0,1,\ldots$, which converges to a local optimal solution of Problem JLCM as $\beta$ grows sufficiently large.
\end{theorem}

\vspace{0.03in}
\noindent{\bf Remark:} To prove Theorem~\ref{th:thm_2}, we show that Algorithm JLCM generates a series of decreasing objective values $z+\sum_jF(\Lambda_j)+\theta\hat{C}$ of Problem JLCM with a modified cost function:
\begin{eqnarray}
\hat{C} = \sum_{i=1}^r \sum_{j=1}^m V_j \frac{\log(\beta\pi_{i,j}+1 )}{\log\beta }. \label{eq:modified_obj}
\end{eqnarray}
The key idea in our proof is that the linear approximation of storage cost function in (\ref{eq:app}) can be seen as a sub-gradient of $V_j{\log(\beta\pi_{i,j}+1 )}/ {\log\beta}$, which converges to the real storage cost function as $\beta \rightarrow \infty$, i.e.,
\begin{eqnarray}
\lim_{\beta \rightarrow \infty}  V_j \frac{\log(\beta\pi_{i,j}+1 )}{\log\beta } = V_j {\bf 1}_{(\pi_{i,j}>0)}.
\end{eqnarray}
Therefore, a converging sequence for the modified objective $z+\sum_jF(\Lambda_j)+\theta\hat{C}$ also minimizes Problem JLCM, and the optimization gap becomes zero as $\beta \rightarrow \infty$. Further, it is shown that $\hat{h}$ is a concave function. Thus, minimizing $z+\sum_jF(\Lambda_j)+\theta\hat{h}$ can be viewed as optimizing the difference between 2 convex objectives, namely $z+\sum_jF(\Lambda_j)$ and $-\theta\hat{h}$, which can be also solved via a Difference-of-Convex Programming (DCP). In this context, our linear approximation of cost function in (\ref{eq:app}) can be viewed as an approximated supper-gradient in DCP. Please refer to \cite{DCP:12}  for a comprehensive study of regularization techniques in DCP to speed up the convergence of Algorithm JLCM.

%% file: 6_impl_latest.tex
\section{Implementation and Evaluation}

\label{sec:impl}
\vspace{-.1in}

\input{6_tahoe}

\vspace{-.15in}

\subsection{Experiments and Evaluation}

{\bf Validate our latency analysis.} While our service delay bound applies to arbitrary distribution and works for systems hosting any number of files, we first run an experiment to understand actual service time distribution on our testbed. We upload a 50MB file using a $(7,4)$ erasure code and measure the chunk service time. Figure \ref{fig:service_dist} depicts the Cumulative Distribution Function (CDF) of the chunk service time. Using the measured results, we get the mean service time of $13.9$ seconds with a standard deviation of 4.3 seconds, second moment of 211.8 $s^2$ and the third moment of 3476.8 $s^3$. We compare the distribution to the exponential distribution( with the same mean and the same variance, respectively) and note that the two do not match. It verifies that actual service time does not follow an exponential distribution, and therefore, the assumption of exponential service time in \cite{ISIT:12,MG1:12} is falsified by empirical data. The observation is also evident because a distribution never has positive probability for very small service time. Further, the mean and the standard deviation are very different from each other and cannot be matched by any exponential distribution.

\begin{figure*}[!th]
\vspace{-1mm}
\begin{minipage}{0.47\textwidth}
\begin{center}
\scalebox{0.35}{\includegraphics[draft=false]{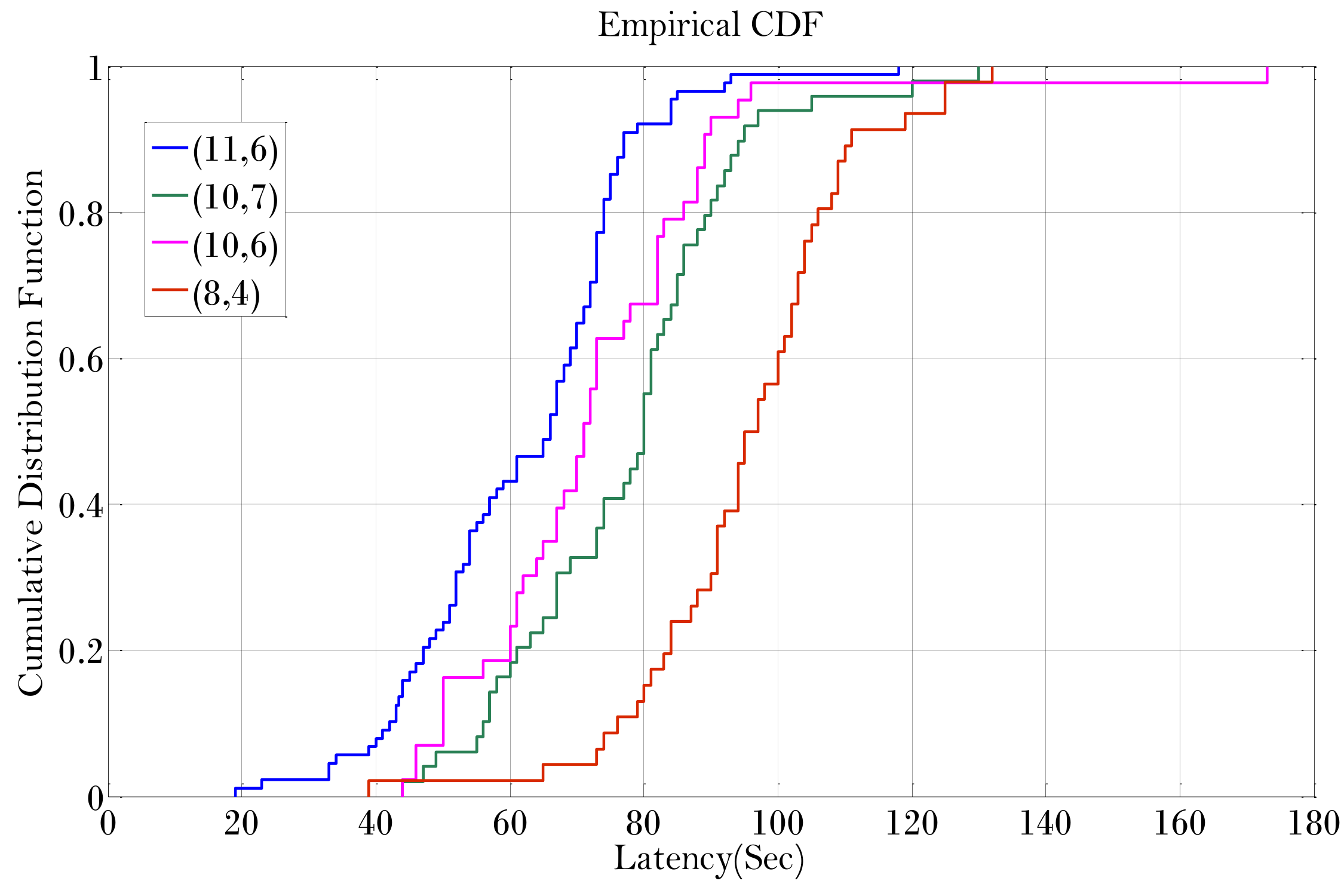}}
\vspace{2mm}
\caption{Actual service latency distribution of an optimal solution from Algorithm JLCM for $1000$ files of size $150MB$ using erasure code (11,6), (10,7), (10,6) and (8,4) for each quarter with aggregate request arrival rates are set to $\lambda_i=0.118$ /sec}
\label{fig:latency_distribution}
\end{center}
\end{minipage}
\begin{minipage}{0.47\textwidth}
\begin{center}
\scalebox{0.35}{\includegraphics[draft=false]{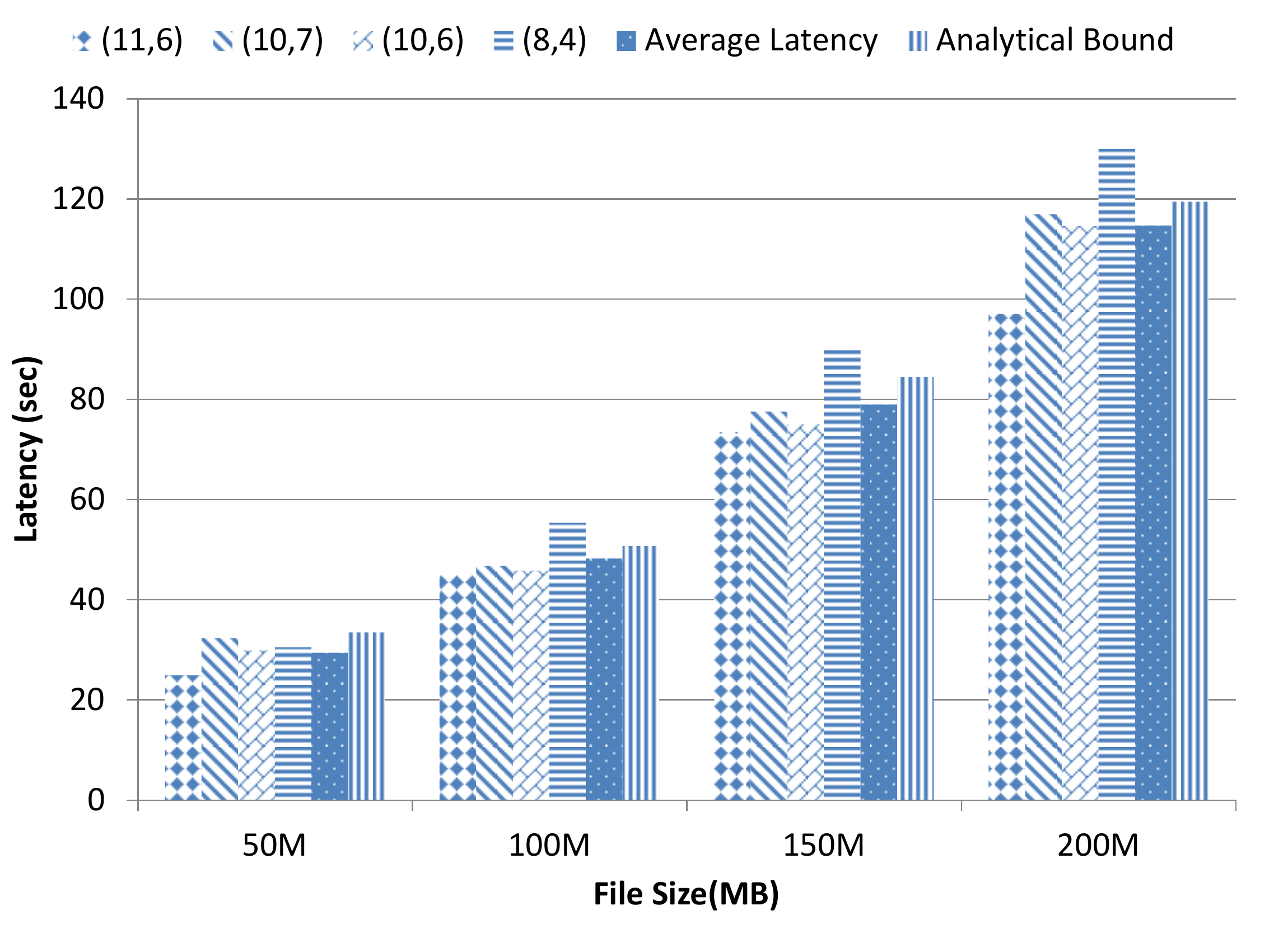}}
\vspace{-4mm}
\caption{Evaluation of different chunk sizes. Latency increases super-linearly as file size grows due to queuing delay. Our analytical latency bound taking both network and queuing delay into account tightly follows actual service latency.}
\label{fig:varying_size}
\end{center}
\end{minipage}
%
%

\begin{minipage}{0.47\textwidth}
\begin{center}
\vspace{2mm}
\scalebox{0.32}{\includegraphics[draft=false]{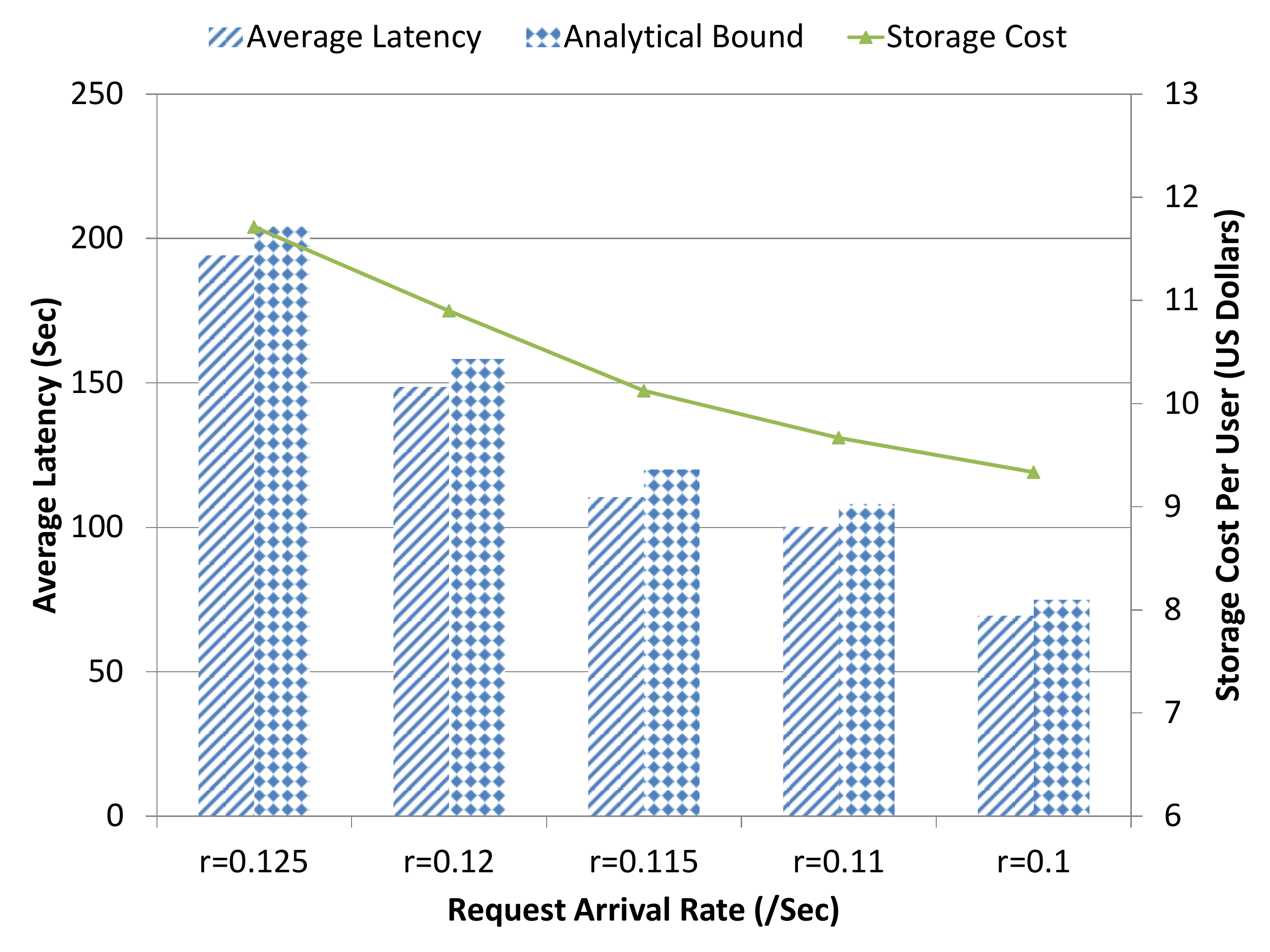}}
\vspace{-2mm}
\caption{Evaluation of different request arrival rates. As arrival rates increase, latency increases and becomes more dominating in the latency-plus-cost objective than storage cost. The optimal solution from Algorithm JLCM allows higher storage cost, resulting in a nearly-linear growth of average latency. }
\label{fig:workload}
\end{center}
\end{minipage}
\hspace{0.2cm}
\begin{minipage}{0.47\textwidth}
\begin{center}
\scalebox{0.59}{\includegraphics[draft=false]{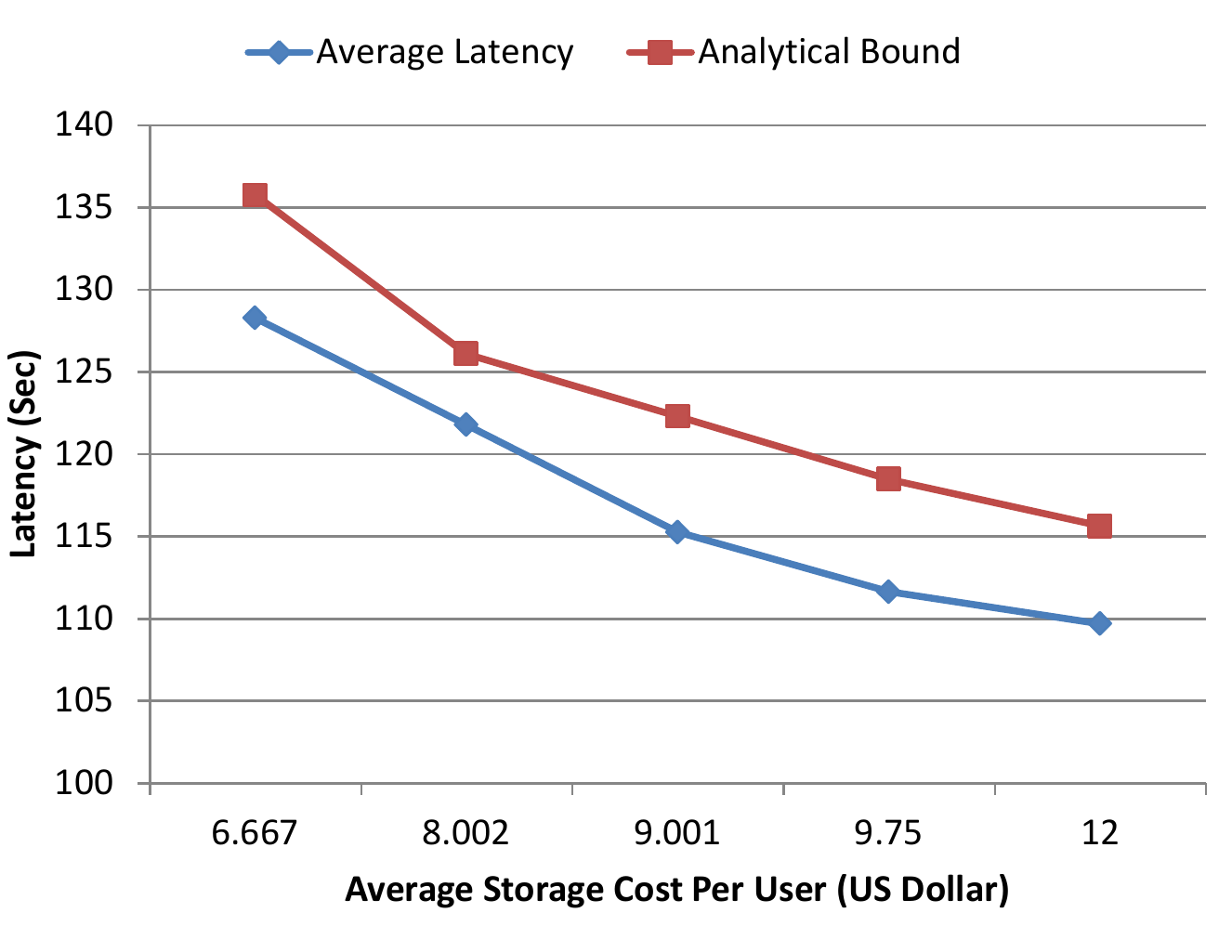}}
\vspace{-4mm}
\caption{Visualization of latency and cost tradeoff for varying $\theta=0.5$ second/dollar to $\theta=200$ second/dollar. As $\theta$ increases, higher weight is placed on the storage cost component of the latency-plus-cost objective, leading to less file chunks and higher latency. }
\label{fig:tradeoff}
\end{center}
\end{minipage}
\vspace{-.3in}
\end{figure*}

Using the service time distribution obtained above, we compare the upper bound on latency that we propose in this paper with the outer bound in \cite{Joshi:13}. Even though our upper bound holds for multiple heterogeneous files, and includes connection delay, we restrict our comparison to the case for a single file/homogeneous-file(multiple homogeneous files with exactly the same properties can be reduced to the case of single file) without any connection delay for a fair comparison (since the upper bound in \cite{Joshi:13} only works for the case of a single file/homogeneous files). We plot the latency upper bound that we give in this paper and the upper bound in [Theorem 3, \cite{Joshi:13}] in Figure \ref{fig:latency_compare}. In our probabilistic scheduling, access requests are dispatched uniformly to all storage nodes. We find that our bound significantly outperforms the upper bound in \cite{Joshi:13} for a wide range of $1/\lambda <32$, which represents medium to high traffic regime. In particular, our bound works fine in high traffic regime with $1/\lambda <18$, whereas the upper bound in \cite{Joshi:13} goes to infinity and thus fail to offer any useful information. Under low traffic, the two bounds get very close to each other with a less than $4\%$ gap.

{\bf Validate Algorithm JLCM and joint optimization.} We implemented Algorithm JLCM and used MOSEK \cite{MOSEK}, a commercial optimization solver, to realize the projected gradient routine. For 12 distributed storage nodes in our testbed, Figure~\ref{fig:convergence} demonstrates the convergence of Algorithm JLCM, which optimizes latency-plus-cost over three dimensions: erasure code length $n_i$, chunk placement $\mathcal{S}_i$, and load balancing $\pi_{i,j}$. Convergence of Algorithm JLCM is guaranteed by Theorem~\ref{th:thm_2}. To speed up its calculation, in this experiment we merge different updates, including the linear approximation, the latency bound minimization, and the projected gradient update, into one single loop. By performing these updates on the same time-scale, our Algorithm JLCM efficiently solves the joint optimization of problem size $r=1000$ files.  It is observed that the normalized objective (i.e., latency-plus-cost normalized by the minimum) converges within 250 iterations for a tolerance $\epsilon=0.01$. To achieve dynamic file management, our optimization algorithm can be executed repeatedly upon file arrivals and departures.

To demonstrate the joint latency-plus-cost optimization of Algorithm JLCM, we compare its solution with three oblivious schemes, each of which minimize latency-plus-cost over only a subset of the 3 dimensions: load-balancing (LB), chunk placement (CP), and erasure code (EC). We run Algorithm JLCM for $r=1000$ files of size $150MB$ on our testbed, with $V_j=\$1$ for every $25MB$ storage and a tradeoff factor of $\theta=200$ sec/dollar. The result is shown in Figure.~\ref{fig:optimality}. First, even with the optimal erasure code and chunk placement (which means the same storage cost as the optimal solution from Algorithm JLCM), higher latency is observed in {\em Oblivious LB}, which schedules chunk requests according to a load-balancing heuristic that selects storage nodes with probabilities proportional to their service rates. Second, we keep optimal erasure codes and employ a random chunk placement algorithm, referred to as {\em Random CP}, which adopts the best outcome of 100 random runs. Large latency increment resulted by Random CP highlights the importance of joint chunk placement and load balancing in reducing service latency. Finally, {\em Maximum EC} uses maximum possible erasure code $n=m$ and selects all nodes for chunk placement. Although its latency is comparable to the optimal solution from Algorithm JLCM, higher storage cost is observed. We verify that minimum latency-plus-cost can only be achieved by jointly optimizing over all 3 dimensions.

{\bf Evaluate the performance of our solution.} First, we choose $r=1000$ files of size $150MB$ and the same storage cost and tradeoff factor as in the previous experiment. Request arrival rates are set to $\lambda_i=1.25/(10000 sec)$, for $i=1,4,7,\cdots 997$, $\lambda_i=1.25/(10000 sec)$, for $i=2,5,8,\cdots 998$ and $\lambda_i=1.25/(12000 sec)$, for $i=3,6,9,\cdots 999, 1000$ respectively, which leads to an aggregate file request arrival rate of $\lambda=0.118$ /sec. We obtain the service time statistics (including mean, variance, second and third moment) at all storage nodes and run Algorithm JLCM to generate an optimal latency-plus-cost solution, which results in four different sets of optimal erasure code (11,6), (10,7), (10,6) and (9,4) for each quarter of the 1000 files respectively, as well as associated chunk placement and load-balancing probabilities. Implementing this solution on our testbed, we retrieve the 1000 files at the designated request arrival rate and plot the CDF of download latency for each file in Figure~\ref{fig:latency_distribution}. We note that 95\% of download requests for files with erasure code (10,7) complete within 100 seconds, while the same percentage of requests for files using (11,6) erasure code complete within 32 seconds due to higher level of redundancy. In this experiment erasure code (10,6) outperforms (8,4) in latency though they have the same level of redundancy because the latter has larger chunk size when file size are set to be the same.

To demonstrate the effectiveness of our joint optimization, we vary file size in the experiment from 50MB to 200MB and plot the average download latency of the 1000 individual files, out of which each quarter is using a distinct erasure code (11,6), (10,7), (10,6) and (9,4), and our analytical latency upper bound in Figure~\ref{fig:varying_size} . We see that latency increases super-linearly as file size grows, since it generates higher load on the storage system, causing larger queuing latency (which is super-linear according to our analysis). Further, smaller files always have lower latency because it is less costly to achieve higher redundancy for these files. We also observe that our analytical latency bound tightly follows actual average service latency. 

Next, we varied aggregate file request arrival rate from $\lambda_i=0.125$ /sec to $\lambda_i=0.1$ /sec (with individual arrival rates also varies accordingly), while keeping tradeoff factor at $\theta=2$ sec/dollar and file size at $200MB$. Actual service delay and our analytical bound for each scenario is shown by a bar plot in Figure~\ref{fig:workload} and associated storage cost by a curve plot. Our analytical bound provides a close estimate of service latency. As arrival rates increase, latency increases and becomes more dominating in the latency-plus-cost objective than storage cost. Thus, the marginal benefit of adding more chunks (i.e., redundancy) eventually outweighs higher storage cost introduced at the same time. Figure~\ref{fig:workload} shows that to achieve a minimization of the latency-plus-cost objective, the optimal solution from Algorithm JLCM allows higher storage cost for larger arrival rates, resulting in a nearly-linear growth of average latency as the request arrival rates increase. For instance, Algorithm JLCM chooses (12,6), (12,7), (11,6) and (11,4) erasure codes at the largest arrival rates, while (10,6), (10,7), (8,6) and (8,4) codes are selected at the smallest arrival rates in this experiment. We believe that this ability to autonomously manage latency and storage cost for latency-plus-cost minimization under different workload is crucial for practical distributed storage systems relying on erasure coding.

{\bf Visualize latency and cost tradeoff.} Finally, we demonstrate the tradeoff between latency and storage cost in our joint optimization framework. Varying the tradeoff factor in Algorithm JLCM from $\theta=0.5$ sec/dollar to $\theta=200$ sec/dollar for fixed file size of $200MB$ and aggregate arrival rates $\lambda_i=0.125$ /sec, we obtain a sequence of solutions, minimizing different latency-plus-cost objectives. As $\theta$ increases, higher weight is placed on the storage cost component of the  latency-plus-cost objective, leading to less file chunks in the storage system and higher latency. This tradeoff is visualized in Figure~\ref{fig:tradeoff}. When $\theta=0.5$, the optimal solution from Algorithm JLCM chooses three sets of erasure codes (12,6), (12,7), and (12,4) erasure codes, which is the maximum erasure code length in our framework and leads to highest storage cost (i.e., 12 dollars for each user), yet lowest latency (i.e., 110 sec). On the other hand, $\theta=200$ results in the choice of (6,6),	(8,7), and (6,4) erasure code, which is almost the minimum possible cost for storing the three file, with the highest latency of $128$ seconds. Further, the theoretical tradeoff calculated by our analytical bound and Algorithm JLCM is very close to the actual measurement on our testbed. To the best our our knowledge, this is the first work proposing a joint optimization algorithm to exploit such tradeoff in an erasure-coded, distributed storage system.

%% file: 6_tahoe.tex
\subsection{Tahoe Testbed}

To validate our proposed algorithms for joint latency and cost optimization (i.e., Algorithm JLCM) and evaluate their performance, we implemented the algorithms in {\em Tahoe} \cite{Tahoe}, which is an open-source, distributed filesystem based on the {\em zfec} erasure coding library. It provides three special instances of a generic {\em node}: (a)  {\em Tahoe Introducer}: it keeps track of a collection of storage servers and clients and introduces them to each other.   (b) {\em Tahoe Storage Server}: it exposes attached storage to external clients and stores erasure-coded shares.  (c) {\em Tahoe Client}: it processes upload/download requests and connects to storage servers through a Web-based REST API and the Tahoe-LAFS (Least-Authority File System) storage protocol over SSL.

\begin{figure}[!thbp]
\begin{center}
\scalebox{0.3}{\includegraphics[draft=false]{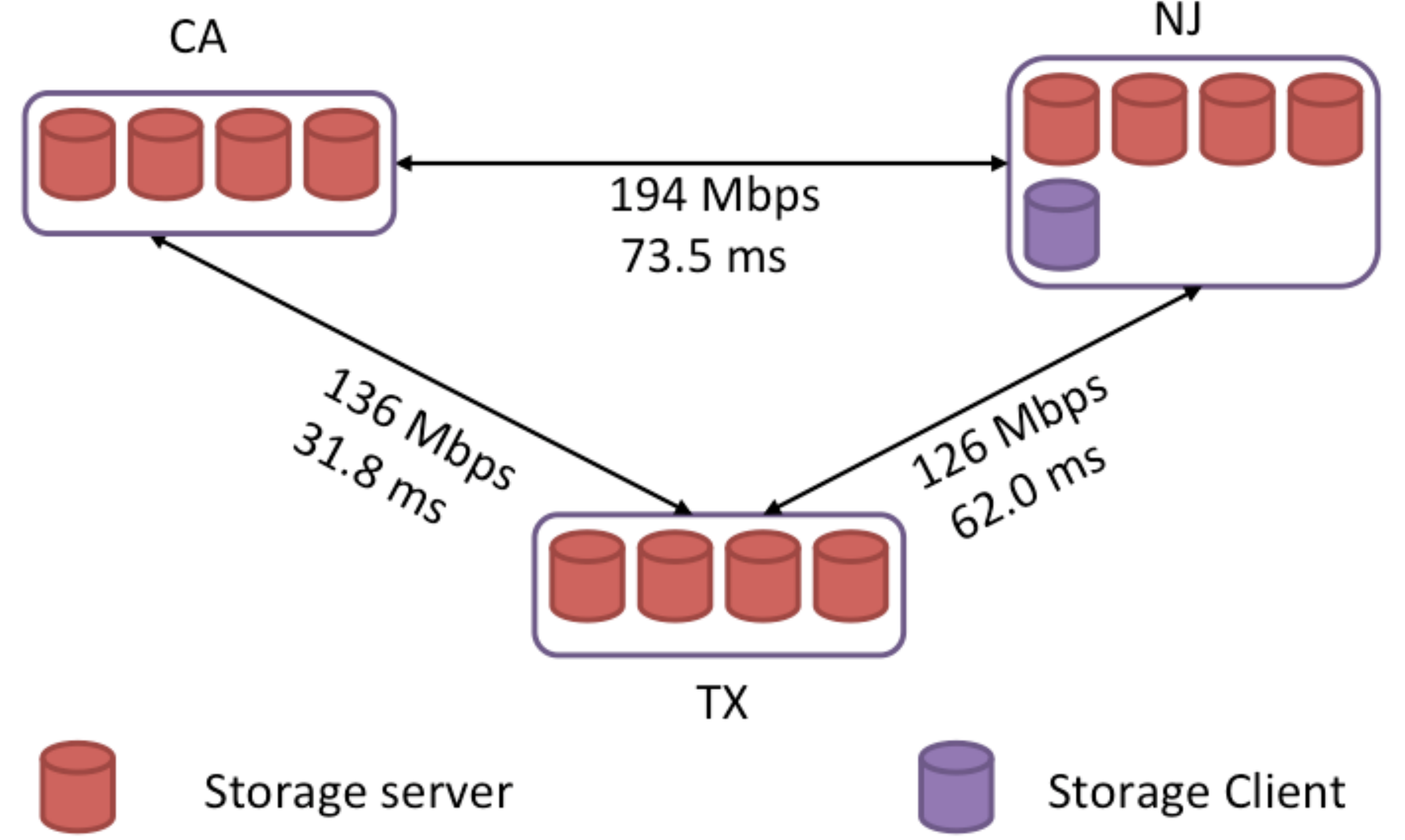}}
\caption{Our Tahoe testbed with average ping (RTT) and bandwidth measurements among three data centers in New Jersey, Texas, and California}
\label{fig:sl-testbed}
\end{center}
\vspace{-.2in}
\end{figure}

Our algorithm requires customized erasure code, chunk placement, and server selection algorithms.  While Tahoe uses a default $(10,3)$ erasure code, it supports arbitrary erasure code specification statically through a configuration file. In Tahoe, each file is
 encrypted, and is then broken into a set of segments, where each segment consists of $k$ blocks.  Each segment is then erasure-coded to produce $n$ blocks (using a $(n,k)$ encoding scheme) and then distributed to (ideally) $n$ distinct storage servers. The set of blocks on each storage server constitute a chunk. Thus, the file equivalently consists of $k$ chunks which are encoded into $n$ chunks and each chunk consists of multiple blocks\footnote{If there are not enough servers, Tahoe will store multiple chunks on one sever. Also, the term ``chunk'' we used in this paper is equivalent to the term ``share'' in Tahoe terminology. The number of blocks in each chunk is equivalent to the number of segments in each file.}. For chunk placement, the Tahoe client randomly selects a set of available storage servers with enough storage space to store $n$ chunks. For server selection during file retrievals, the client first asks all known servers for the storage chunks they might have. Once it knows where to find the needed k chunks (from the k servers that responds the fastest), it downloads at least the first segment from those servers. This means that it tends to download chunks from the ``fastest'' servers purely based on round-trip times (RTT). In our proposed JLCM algorithm, we consider RTT plus expected queuing delay and transfer delay as a measure of latency.

\begin{figure*}[!th]
\begin{minipage}{0.48\textwidth}
\begin{center}
\scalebox{0.52}{\includegraphics[draft=false, trim = 1.25in 3.1in 1.55in 3.4in, clip]{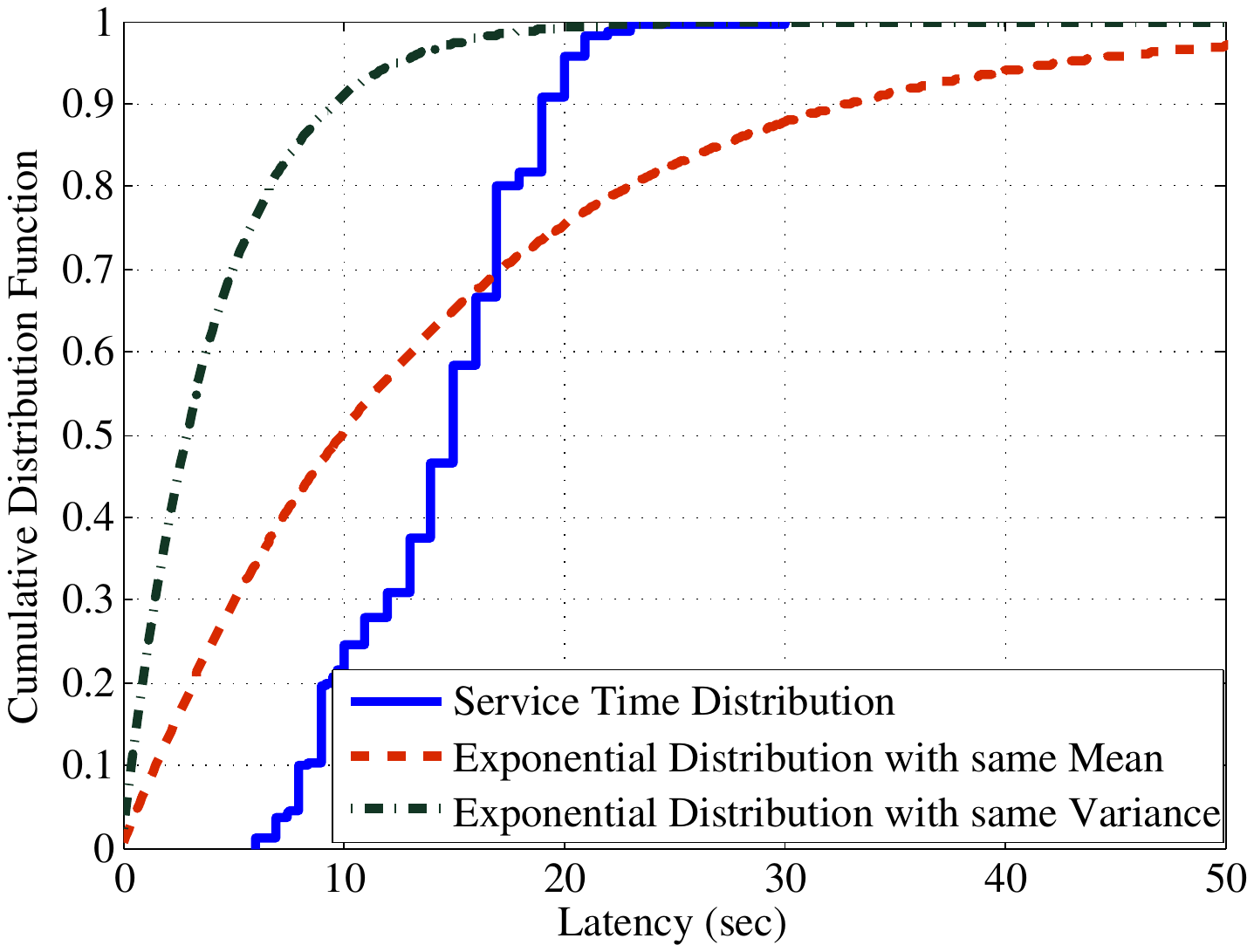}}
\caption{Comparison of actual service time distribution and an exponential distribution with the same mean. It verifies that actual service time does not follow an exponential distribution, falsifying the assumption in previous work \cite{ISIT:12,MG1:12}. }
\label{fig:service_dist}
\end{center}
\end{minipage}
\hspace{0.2cm}
\begin{minipage}{0.48\textwidth}
\begin{center}
\scalebox{0.57}{\includegraphics[draft=false, trim = 1.5in 3.35in 1.8in 3.55in, clip]{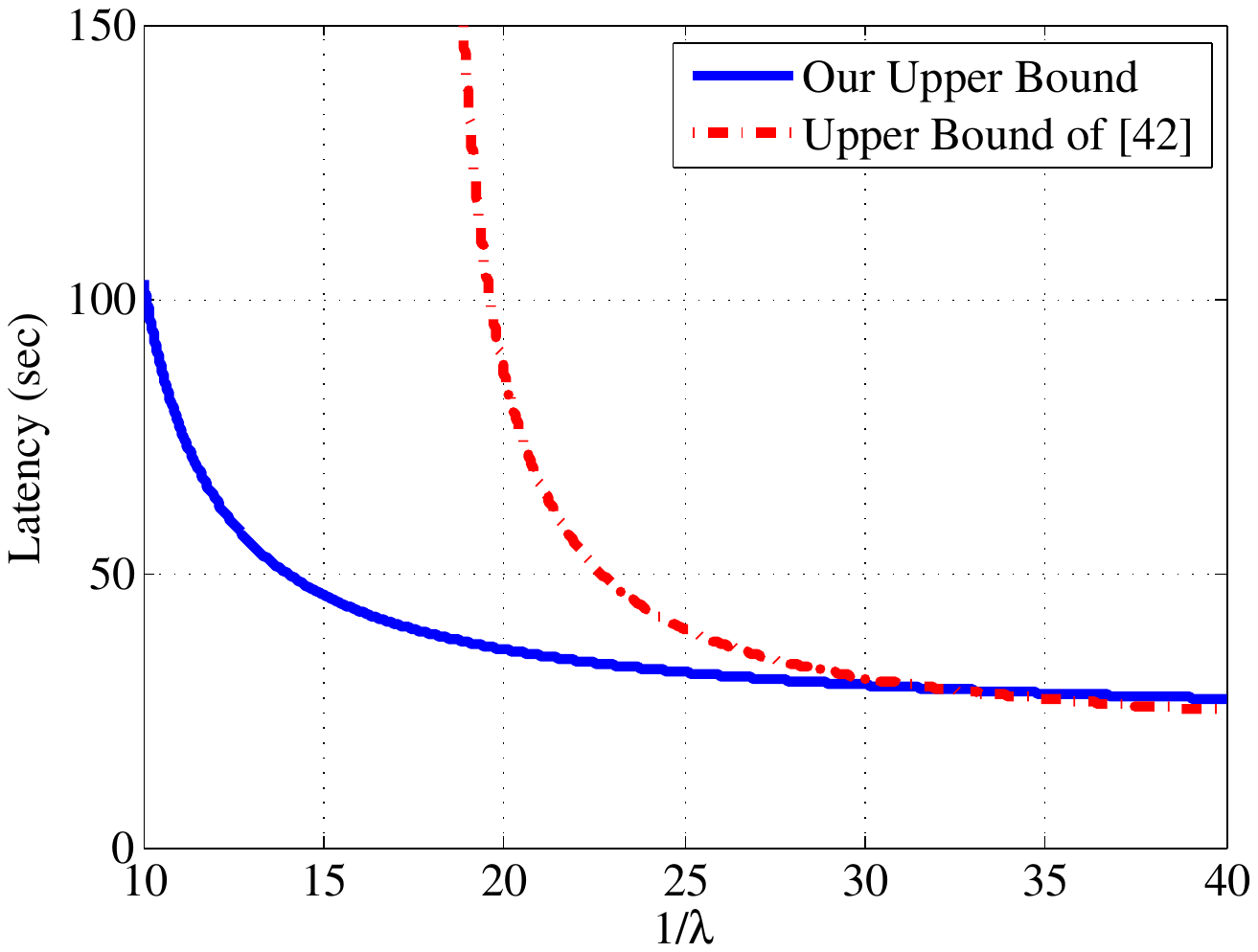}}
\caption{Comparison of our latency upper bound with previous work \cite{Joshi:13}. Our bound significantly improves previous result under medium to high traffic and comes very close to that of \cite{Joshi:13} under low traffic (with  less than $4\%$ gap). }
\label{fig:latency_compare}
\end{center}
\end{minipage}
%
%
\begin{minipage}{0.48\textwidth}
\begin{center}
\includegraphics[scale=.15]{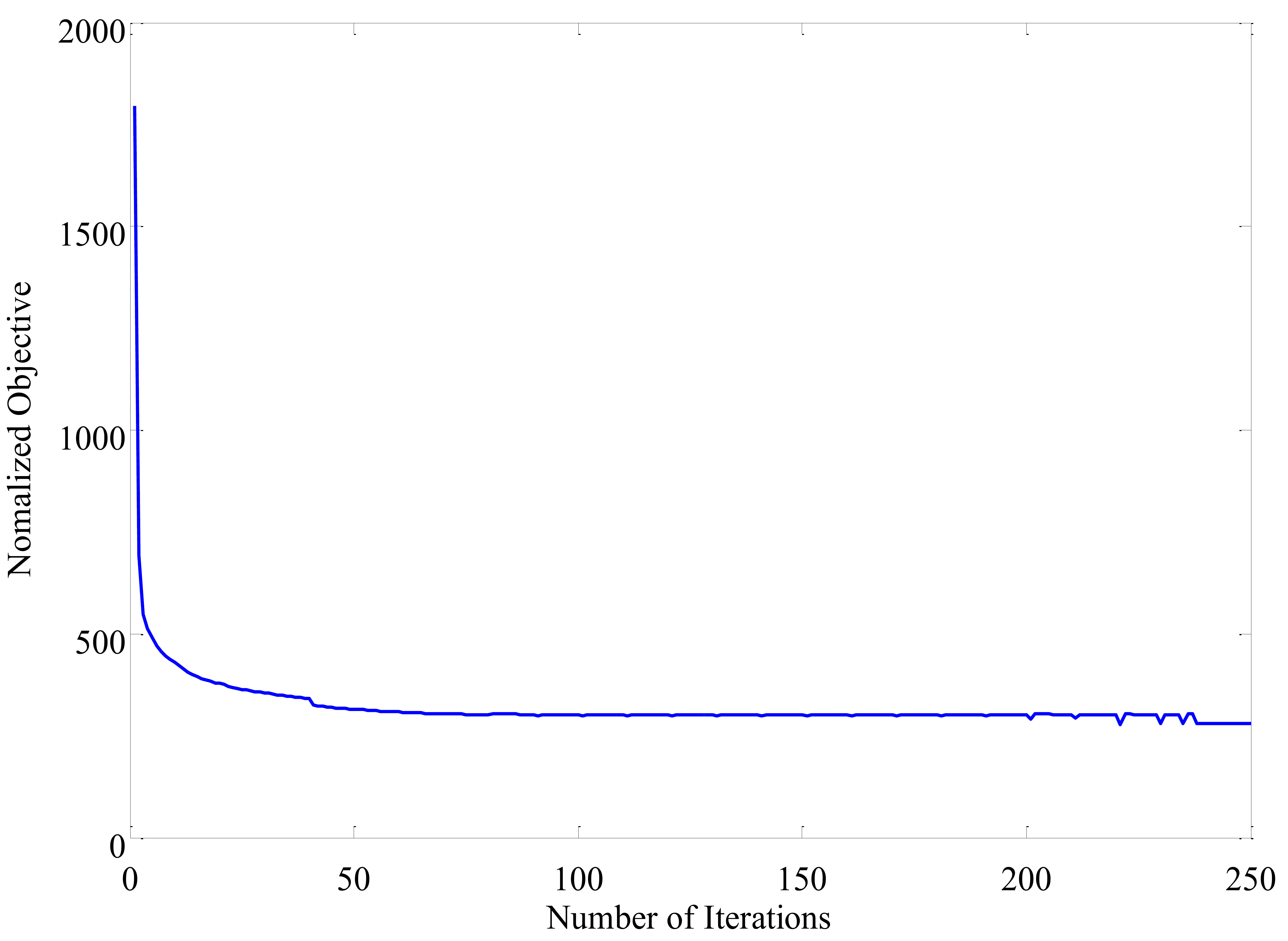}
\vspace{-2mm}
\caption{Convergence of Algorithm JLCM for different problem size with $r=1000$ files for our 12-node testbed. The algorithm efficiently computes a solution in less than 250 iterations.}\label{fig:convergence}
\end{center}
\end{minipage}
\hspace{0.2cm}
\begin{minipage}{0.48\textwidth}
\begin{center}
\includegraphics[scale=.43]{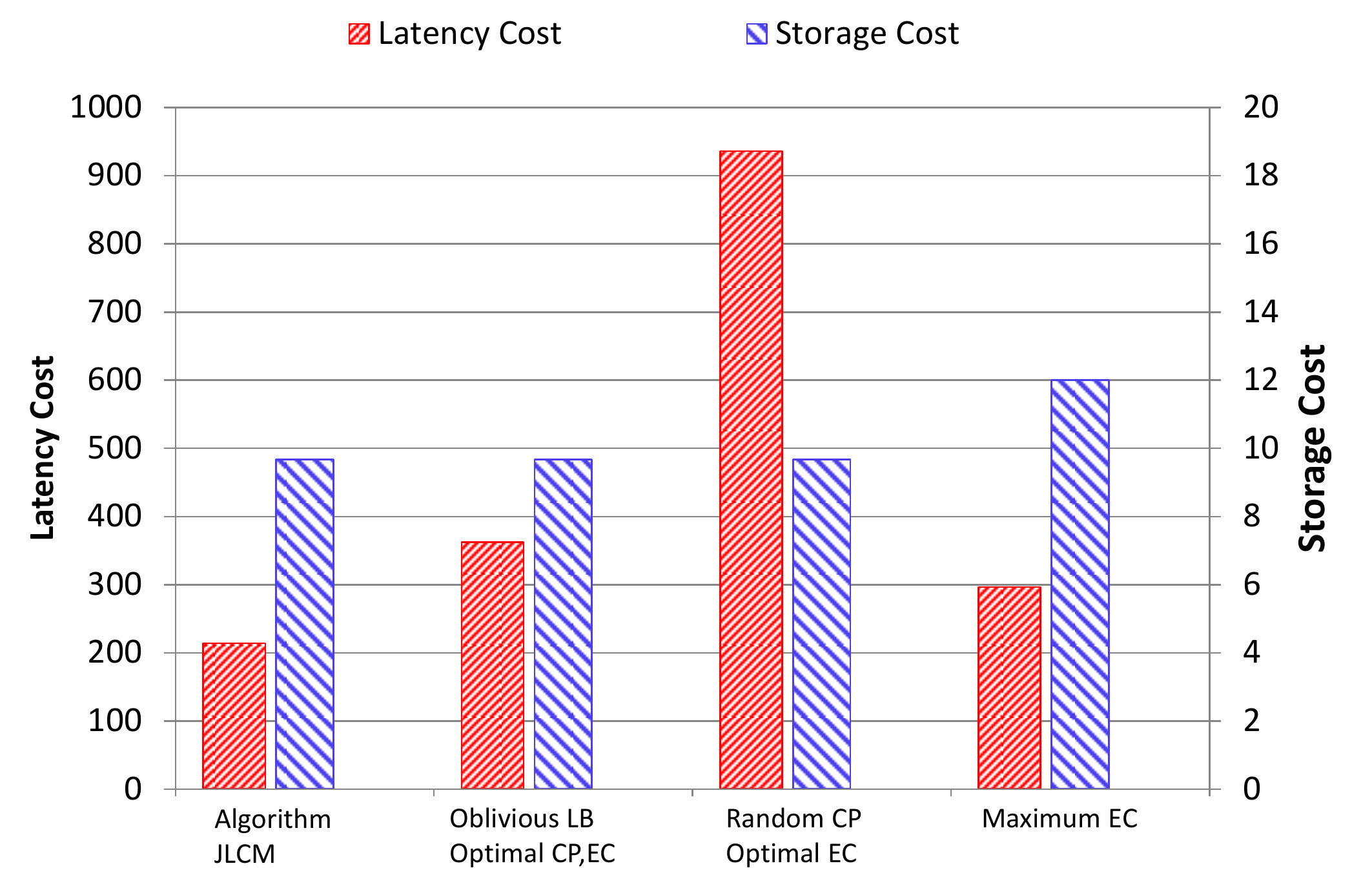}
\caption{Comparison of Algorithm JLCM with some oblivious approaches. Algorithm JLCM minimizes latency-plus-cost over 3 dimensions: load-balancing (LB), chunk placement (CP), and erasure code (EC), while any optimizations over a subset of the dimensions is non-optimal.}\label{fig:optimality}
\end{center}
\end{minipage}
\vspace{-.3in}
\end{figure*}

In our experiment, we modified the upload and download modules in the Tahoe storage server and client to allow for customized and explicit server selection, which is specified in the configuration file that is read by the client when it starts.   In addition, Tahoe performance suffers from its single-threaded design on the client side for which we had to use multiple clients with separate ports to improve parallelism and bandwidth usage during our experiments.

We deployed 12 Tahoe storage servers as virtual machines in an OpenStack-based data center environment distributed in New Jersey (NJ), Texas (TX), and California (CA).  Each site has four storage servers.  One additional storage client was deployed in the NJ data center to issue storage requests.  The deployment is shown in Figure~\ref{fig:sl-testbed} with average ping (round-trip time) and bandwidth measurements listed among the three data centers.  We note that while the distance between CA and NJ is greater than that of TX and NJ, the maximum bandwidth is higher in the former case. The RTT time measured by ping does not necessarily correlate to the bandwidth number. Further, the current implementation of Tahoe does not use up the maximum available bandwidth, even with our multi-port revision.
